\documentclass[twocolumn]{aastex63}
\usepackage{natbib}
\usepackage{color}
\usepackage{mathtools}
\usepackage{hyperref} 

\usepackage{longtable}
\usepackage{cancel,soul,ulem,amsmath} 

\def\hi{H{\sc i}}

\def\arcmin{\hbox{$^{\prime}$}}

\def\cm2{cm$^{-2}$}
\def\cc{cm$^{-3}$}
\def\kms{km s$^{-1}$}
\def\s{s$^{-1}$}
\def\nh3{NH$_3$}
\def\n2h{N$_2$H$^+$}
\def\co{$^{12}$CO}
\def\13co{$^{13}$CO}
\def\c18o{C$^{18}$O}
\def\hc3n{HC$_3$N}
\def\h2{H$_2$}
\def\nh{n(H$_2$)}
\def\c2{[C\,{\sc ii}]}

\def\lc{\>\> ,}

\def\xoh{$X$(OH)}

\shorttitle{OH Survey toward  Dark Clouds}
\shortauthors{Tang et al.}
\graphicspath{{./}{./fig/}}

\begin{document}

\title{OH Evolution in Molecular Clouds }

\correspondingauthor{Ningyu Tang, Di Li}
\email{nytang@nao.cas.cn, dili@nao.cas.cn}

\author[0000-0002-2169-0472]{Ningyu Tang}
\affiliation{CAS Key Laboratory of FAST, National Astronomical Observatories, Chinese Academy of Sciences, Beijing 100101, People's Republic of China}

\author[0000-0003-3010-7661]{Di Li}
\affiliation{CAS Key Laboratory of FAST, National Astronomical Observatories, Chinese Academy of Sciences, Beijing 100101, People's Republic of China}
\affiliation{University of Chinese Academy of Sciences, Beijing 100049, People’s Republic of China}
\affiliation{NAOC-UKZN Computational Astrophysics Centre, University of KwaZulu-Natal, Durban 4000, South Africa}

\author[0000-0003-0355-6875]{Nannan Yue}
\affiliation{CAS Key Laboratory of FAST, National Astronomical Observatories, Chinese Academy of Sciences, Beijing 100101, People's Republic of China}
\affiliation{University of Chinese Academy of Sciences, Beijing 100049, People’s Republic of China}

\author[0000-0003-3948-9192]{Pei Zuo}
\affiliation{Kavli Institute for Astronomy and Astrophysics, Peking University, Beijing 100871, China}
\affiliation{International Centre for Radio Astronomy Research (ICRAR), The University of Western Australia, 35 Stirling Hwy, Crawley, WA 6009, Australia}
\affiliation{University of Chinese Academy of Sciences, Beijing 100049, People’s Republic of China}

\author[0000-0002-5286-2564]{Tie Liu}
\affiliation{Shanghai Astronomical Observatory, Chinese Academy of Sciences, 80
Nandan Road, Shanghai 200030, People’s Republic of China}
\affiliation{Key Laboratory for Research in Galaxies and Cosmology, Chinese Academy of Sciences, 80 Nandan Road, Shanghai 200030, People’s Republic of China}

\author[0000-0002-1583-8514]{Gan Luo}
\affiliation{CAS Key Laboratory of FAST, National Astronomical Observatories, Chinese Academy of Sciences, Beijing 100101, People's Republic of China}
\affiliation{University of Chinese Academy of Sciences, Beijing 100049, People’s Republic of China}

\author{Longfei Chen}
\affiliation{CAS Key Laboratory of FAST, National Astronomical Observatories, Chinese Academy of Sciences, Beijing 100101, People's Republic of China}

\author{Sheng-Li Qin}
\affiliation{ Department of Astronomy, Yunnan University, Kunming 650091, China}

\author{Yuefang Wu}
\affiliation{ Department of Astronomy, School of Physics, Peking University, 100871
Beijing, China}
\affiliation{Kavili Institute for Astronomy and Astrophysics, Peking University, 100871
Beijing, China}

\author{ Carl Heiles}
\affiliation{ Department of Astronomy, University of California, Berkeley, 601 Campbell Hall 3411, Berkeley, CA 94720-3411, USA}



\begin{abstract}

We have conducted  OH 18 cm survey  toward  141 molecular clouds in various environments, including 33 optical dark clouds, 98 Planck Galactic cold clumps (PGCCs) and 10 Spitzer dark clouds  with the Arecibo telescope.  The deviations from   local thermal equilibrium are common for intensity ratios of both  OH main lines and satellite lines. Line intensity of OH 1667 MHz  is found to  correlate linearly with visual extinction $A\rm_V$ when $A\rm_V$  is less than 3 mag. It was converted into OH column density by adopting excitation temperature derived from Monte Carlo simulations with one sigma uncertainty. The relationship between OH abundance  $X$(OH) relative to \h2  and  $A\rm_V$ is found to follow an empirical formula, 
 \begin{equation}
\nonumber
\frac{X(\textrm{OH})}{10^{-7}} = 1.3^{+0.4}_{-0.4} + 6.3^{+0.5}_{-0.5}\times \textrm{exp}(-\frac{A_\textrm{V}}{2.9^{+0.6}_{-0.6}}). 
\end{equation}
Linear correlation is found between OH and \13co\ intensity.  Besides,  nonthermal velocity dispersions of OH and  \13co\ are closely correlated.  These results imply tight chemical evolution and spatial occupation between OH and \13co.  No  obvious correlation is found between column density and nonthermal velocity dispersion of OH and \hi\ Narrow Self-Absorption (HINSA), indicating different chemical evolution and spatial volume occupation between OH and HINSA.  Using the age information of HINSA analysis, OH abundance $X$(OH) is found to increase linearly with cloud age, which is consistent with previous simulations. Fourteen OH components without corresponding CO emission were detected, implying the effectiveness of OH in tracing the  `CO-dark' molecular gas.  
\end{abstract}



\section{Introduction} \label{sec:intro}

The hydroxyl radical (OH) transition was firstly detected at 18 cm  band in the interstellar medium (ISM) in 1963 \citep{1963Natur.200..829W}.  It was found to  exist  in various environments,  for instance,   dark molecular clouds \citep[e.g.,][]{1969ApJ...157..123H,1973ApJ...185..857C},  diffuse/translucent molecular clouds \citep[e.g.,][]{1990A&A...240..400G, 2010MNRAS.407.2645B, 2012AJ....144..163C, 2014MNRAS.439.1596D}, and  the regions without CO emission while OH is detected in emission  \citep[e.g.,][]{2012AJ....143...97A,2015AJ....149..123A,2019ApJ...883..158B} or  absorption toward continuum sources \citep[e.g.,][]{2018ApJS..235....1L, 2018A&A...618A.159R}.  

\begin{figure}
\begin{center}  
  \includegraphics[width=0.48\textwidth]{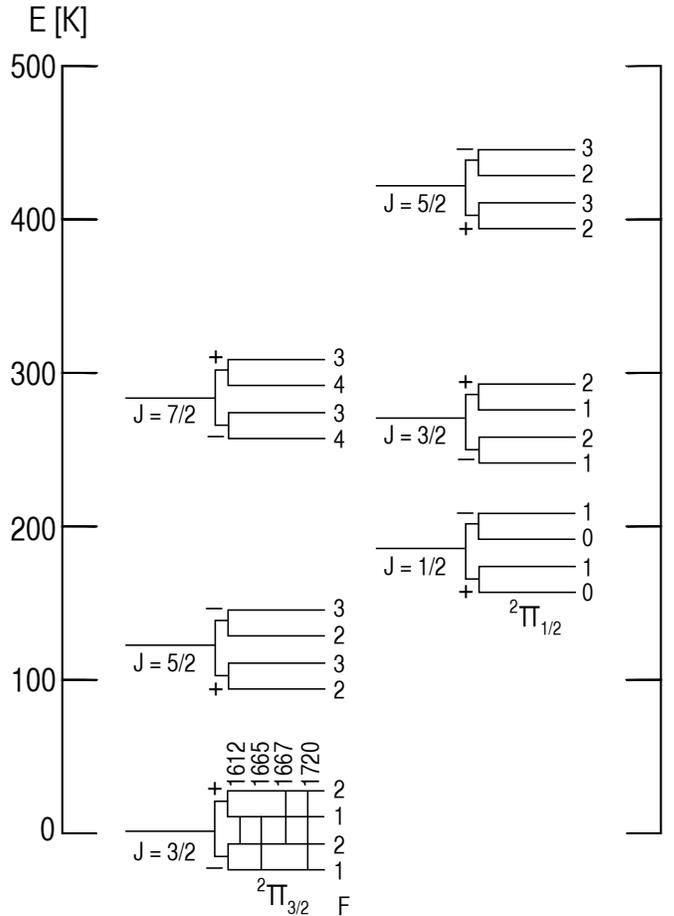}
  \caption{Energy level diagram of OH  $^2\Pi_{3/2}$ and $^2\Pi_{1/2}$ states \citep{2015ApJ...815...13E}. $J$ and $F$ represent total angular momentum quantum number without and with the nuclear spin, respectively. `+' and `-' represent the $\Lambda$-type doubling levels. A transition between two energy levels is allowed when the energy levels have different $\Lambda$-types and satisfy the $\Delta J=0, \pm 1$ and $\Delta F=0, \pm 1$ condition. The OH 1612, 1665, 1667 and 1720 MHz transitions happen between hyperfine levels of the $^2\Pi_{3/2}$ (J=3/2).} 
\label{fig:oh_levels}
\end{center}
\end{figure}

There are four hyperfine transitions (1612.231, 1665.402, 1667.359 and 1720.530 MHz)  of  $^2\Pi_{3/2}$ J=3/2  ground state of OH (Fig. \ref{fig:oh_levels}). The line strengths of these four transitions are considered to follow $T\rm_A^{1612}$:$T\rm_A^{1665}$:$T\rm_A^{1667}$:$T\rm_A^{1720}$= 1:5:9:1 under the local thermodynamic equilibrium (LTE).  It was found that the observed line strengths ratio always deviate from the intrinsic  1:5:9:1 ratio due to the effect of surrounding environment.  In most cases, the 1612 and 1720 MHz  satellite lines  appear as `conjugate' behavior (with one transition being in emission and the other one being in absorption) and  `flip'  behavior (with transition from emission to absorption in 1612 or 1720  transition alone), which may originate from the infrared pumping by its surrounding source \citep[e.g.,][]{1977ApJ...216..308C}. Due to sensitive dependence on radiation field, the OH satellite  lines are ideal indicator of external illumination environment of the clouds. The excitation of main lines (1665 and 1667 MHz) should be less affected because the quantum number $F$ values of upper and lower population levels are same.  However, OH absorptions toward continuum sources indicate that the non-thermal  excitation  are common for main lines \citep[e.g.,][]{1981A&A....98..271D, 2018ApJS..235....1L}.    

OH abundance relative to \h2\ is a critical parameter for studying OH evolution of ISM.  Previous observations suggested a  commensal value of $1 \times 10^{-7}$ toward nearby interstellar medium (ISM) \citep[e.g.,][]{1974A&A....35..445S, 1974ApJ...194..525T, 1976ApJ...209..778C}. \citet{2018ApJ...862...49N} found a consistent result  by explicit OH absorption observations against continuum sources.  Most of the OH abundance measurements were made toward the positions with visual extinction $A_V \leq 5$ mag.  

Previous OH observations focus on nearby giant molecular clouds, e.g., Ophiuchus cloud \citep{1978ApJ...220..864M, 2015ApJ...815...13E} or diffuse region of translucent clouds \citep{2012AJ....144..163C}.  In order to systematically investigate OH evolution in molecular clouds in various environments, we conducted  OH survey toward molecular clouds with different  environments, which  includes optical dark clouds,  Spitzer 8 $\mu$m dark clouds and Planck cold clumps. With a density range of 10$^2$ to 10$^5$  cm$^{-3}$, visual extinction range of 1 to 60 mag and scale size from 0.1 pc to tens of pc, these clouds are undergoing different evolution stages, from star-less quiet state to cluster star formation \citep{2013ApJS..209...37M, 2016ApJS..224...43Z, 2017MNRAS.470.2253N}.  The observations would provide us a full view of the  OH evolution in molecular clouds.

This paper is organized as follows.  In section \ref{sec:observations}, we describe the details of observations and data.  In section \ref{sec:results}, the observation results of OH are described. OH abundance and the relationships between OH and other molecular tracers are investigated in section \ref{sec:analysis} and \ref{sec:hinsa_analysis}. Discussion and summary  are presented in section \ref{sec:discussion} and  section \ref{sec:conclusions}, respectively.

\section{Observations and Archival Data}
\label{sec:observations}

\subsection{Source Selection}
\label{subsec:spatial_distri}

The optical dark clouds (ODCs) are selected based on the catalog of \citet{2002A&A...383..631D}.  Those ODCs with angular diameter that is larger than the beam width of Arecibo telescope ($\sim$ 3.1 arcmin at 1.6 GHz) were selected.  Finally,  data were observed toward 33 ODCs, including 21 sources in the first quarter  and 12 sources in the anti-center direction.  The visual extinction of 29 ODCs ranges from 1 to 5 mag, while visual extinction of 4 sources (LDN621, LDN638, LDN649 and LDN673) exceeds the value of 5 mag. 

The Planck Galactic cold clumps (PGCCs) were selected from the catalog of \citet{2011A&A...536A..23P}.  Those clumps having CO emission \citep{2012ApJ...756...76W} and the feature of \hi\  self absorption were preferred.  A total of 98 PGCCs with visual extinction range of 0.3 to 5 mag  were observed.  Same as ODCs, PGCCs in both first quarter and anti-center direction were selected.

The 10 Spitzer dark clouds (SDCs) with visual extinction A$_V \geq 5$ mag  were selected from the catalog of \citet{2009A&A...505..405P}. All these SDCs were confirmed to have Young Stellar Object (YSO) associated. 

\begin{figure*}
\begin{center}  
  \includegraphics[width=0.95\textwidth]{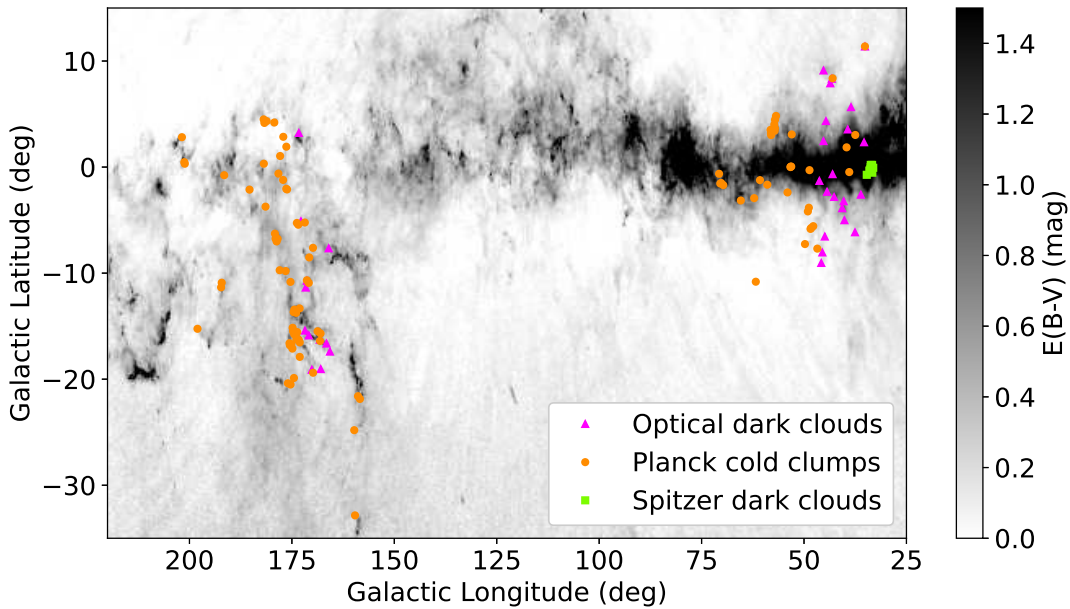}
  \caption{Spatial distribution of sources overlaid on the Planck extinction map. The value of extinction ranges from 0 (white) to 1.5 (black) mag.} 
\label{fig:space-distri}
\end{center}
\end{figure*}

The spatial distribution of these sources is shown in Fig. \ref{fig:space-distri}. These sources are almost symmetric with Galactic latitude in the first quart of Milky Way and associated with giant molecular clouds (e.g., Taurus) in the anti-center direction. Spectral observations and auxiliary data are described as follows.

\subsection{OH Observations with Arecibo Telescope}
\label{subsec:OH}

Observations of the OH 18 cm  transitions  were carried out with the Arecibo 300-m telescope, whose beam size is 3.5 arcmin at 1.6 GHz. Total-ON mode was adopted during observations since the ripple of  spectral bandpass is stable and has a width that is much larger than that of typical Galactic spectral lines ($\sim$ several \kms).  The bandpass after removing  windows containing OH lines was subtracted by linear or polynomial fitting.   We adopted the Interim Correlator backend with a bandwidth of 3.125 MHz  in 8192 channels, leading to a spectral resolution of   0.069 km s$^{-1}$ at 1.66 GHz.  The spectra were smoothed by a factor of two, resulting in spectral resolution of 0.138 km s$^{-1}$.  The observation date,  backend setting  and sensitivity are different for three categories of samples. 

$ODC$:  Three transitions of OH, 1612.2310, 1665.4018, and 1667.3590 MHz were obtained during  September and  October, 2018.  Each source was integrated for 20 minutes, leading to a spectral root-mean-square (rms) of $\sim$21 mK in T$\rm_A$ under velocity resolution of 0.138 km s$^{-1}$ at 1.66 GHz.  

$PGCC$:  Four transitions of OH, 1612.2310, 1665.4018, 1667.3590 and 1720.5300 MHz were obtained between November 2013 and November 2015.  Total  integration time is 5 minutes for each source, resulting in a rms of  $\sim$ 44 mK in T$\rm_A$ under velocity resolution of 0.138 km s$^{-1}$ at 1.66 GHz.

$SDC$:  The observations were taken in November 2013 with same setting as that of PGCCs.

A main beam efficiency of 0.5 \citep{2001PASP..113.1247H} was used to convert the spectra from antenna temperature into main-beam brightness temperature. 

\subsection{Observations of CO Molecular Lines}
\label{subsec:CO}

$^{12}$CO(1-0) and its isotopes ($^{13}$CO(1-0) and C$^{18}$O(1-0)) observations toward 33 ODCs, 11 PGCCs and 10 SDCs   were made with Delingha 13.7m during May and June, 2020. The beam width and main-beam efficiency of the telescope are $\sim$ 50 arcsec  and  $\sim$0.51 at 115 GHz. The system temperatures of these observations vary from 230 K to 350 K, with a typical value of 250 K. The backend has a bandwidth of 1 GHz and channel width of 61 kHz, which corresponds a velocity resolution of 0.16 \kms\ at 115.271 GHz.  Integration time of each source was around 10 minutes to reach a spectral rms of $\sim$ 0.15 K (T$\rm_{mb}$) in 0.16 \kms. 

CO data of rest 87 PGCCs were obtained from CO survey with Delingha 13.7m in 2011 \citep{2012ApJ...756...76W}.  Channel rms of these spectra is  $\sim$ 0.2 K (T$\rm_A$) in 0.16 \kms.  

\subsection{ \hi\ and N{\rm(\h2)} Data}
\label{subsec:hi-av}

\hi\ spectra toward ODCs were obtained simultaneously with OH data, resulting in rms of $\sim$15 mK (T$\rm_A$) in 0.15 km s$^{-1}$. \hi\ data of rest sources were extracted from the Galactic Arecibo L-band Feed Array \hi \citep[GALFA-\hi;][]{2018ApJS..234....2P}, with a rms of 0.33 K in 0.18 km s$^{-1}$. 

The total column density of \h2\ , N(\h2) were derived by dust emission. The values of N(\h2) for PGCCs and SDCs were obtained from \citet{2016A&A...594A..28P} and \citet{2016A&A...590A..72P}. For ODCs, we utilized the  E(B-V) data  from \citet{1998ApJ...500..525S} (SFD98). The conversion factor of 3.1 between visual extinction $A\rm_V$ and E(B-V) was adopted \citep{1975A&A....43..133S}.  N(\h2) were derived by eliminating \hi\ contribution along sightline. 

\section{Results }
\label{sec:results}

\subsection{OH Thermal Emission}
\label{sec:detection}

\begin{figure*}
\gridline{\fig{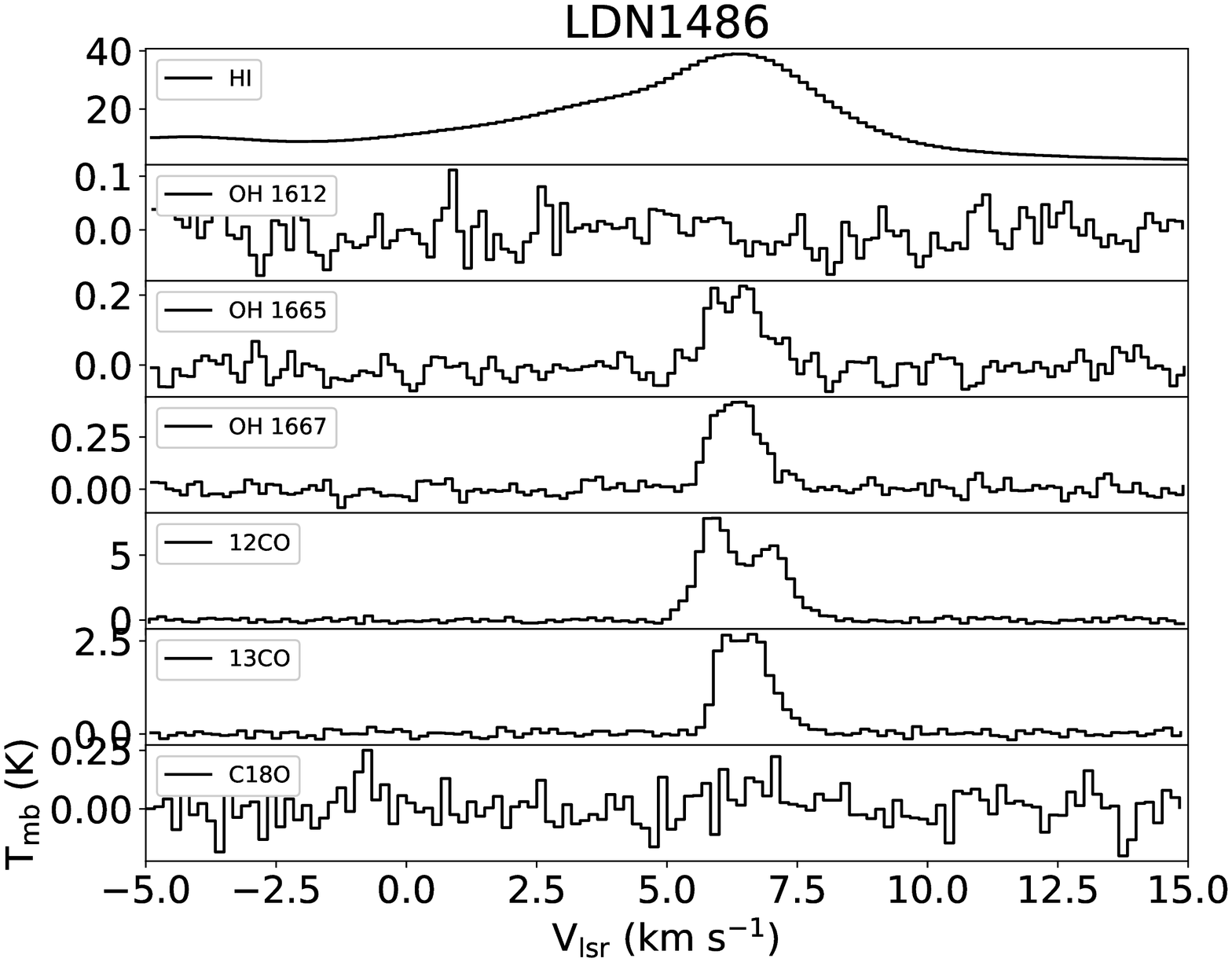}{0.48\textwidth}{(a)}
              \fig{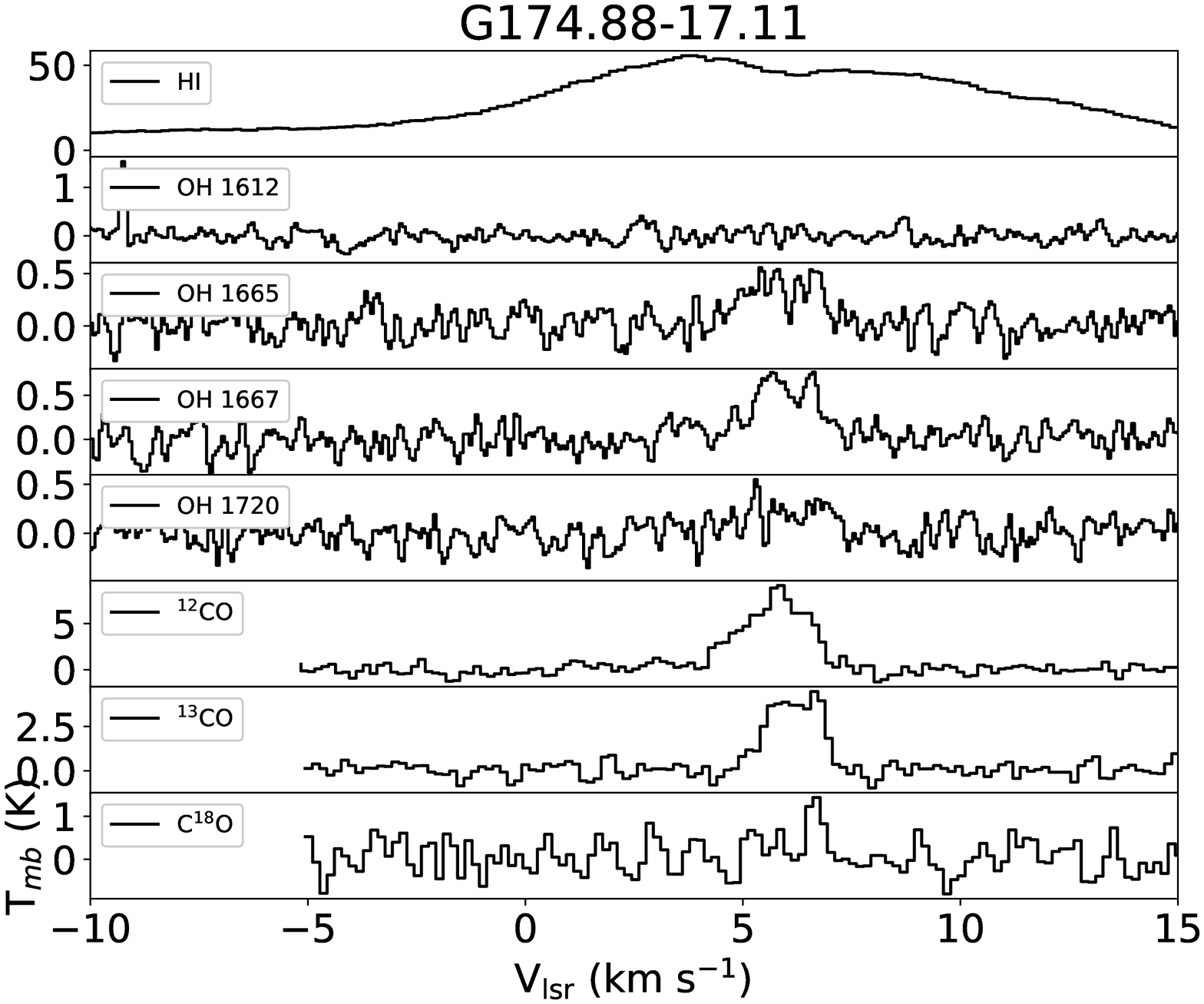}{0.48\textwidth}{(b)}
              }
\gridline{\fig{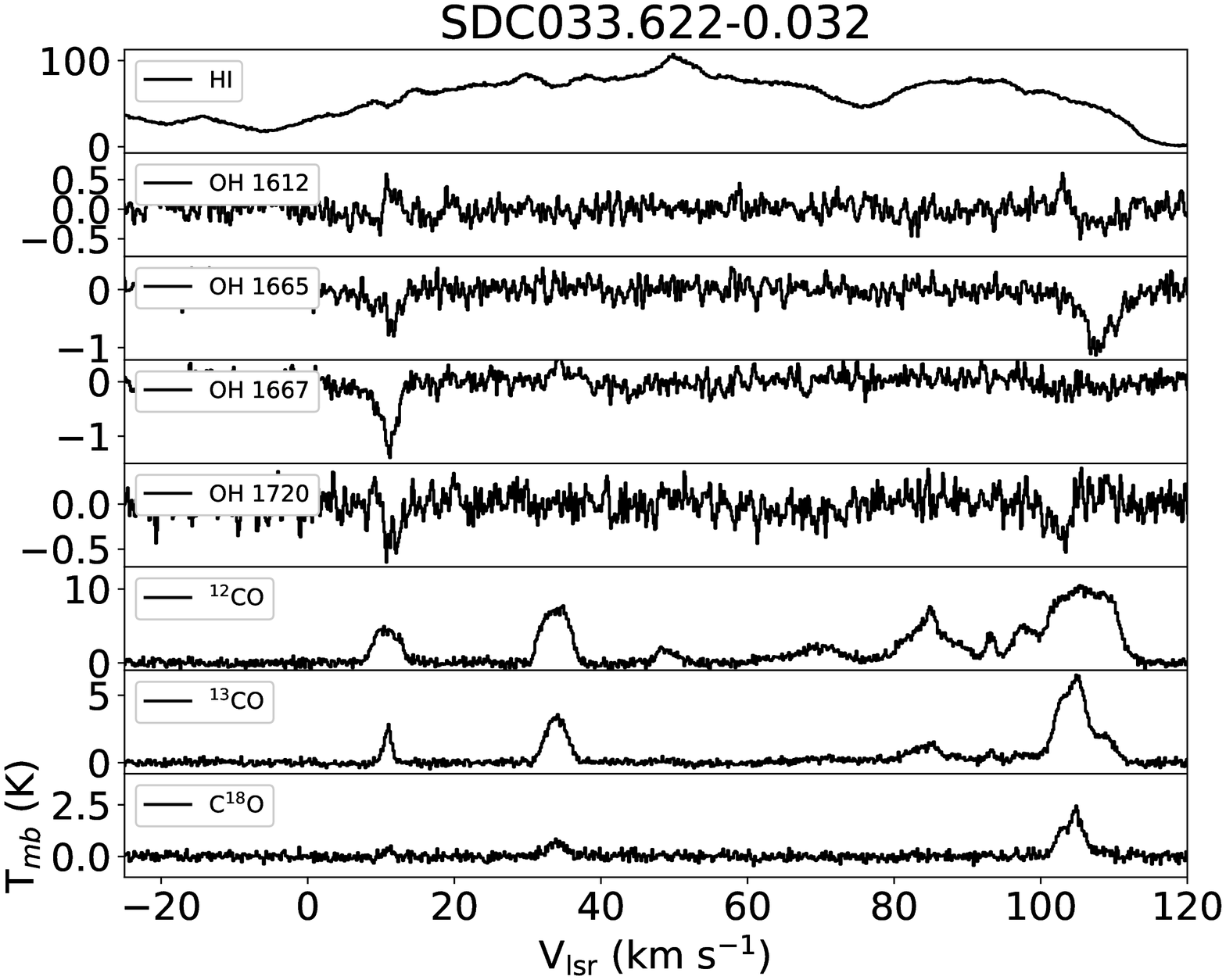}{0.48\textwidth}{(c)}}
\caption{\hi, OH and CO spectra toward LDN1486 (ODCs), G174.88-17.11 (PGCCs) and SDC033.622-0.032 (SDCs).  The continuum levels of these spectra were fitted and subtracted. }
\label{fig:oh_spectra}
\end{figure*}

Sample spectra with OH detection toward ODCs, PGCCs and SDCs are shown in Fig. \ref{fig:oh_spectra}. OH main-lines always appear as emission in the ODCs and PGCCs while appear as absorption in SDCs.  Strong continuum background (e.g., from H II region) associated with  star formation regions may contribute to the absorption feature in SDCs.  We identified OH components in velocity by applying Gaussian decomposition.  The results are shown in Table \ref{table:ohfitresult}.  Most sources have only one OH/CO component in velocity. Multiple OH/CO components with close central velocities were detected toward  33\% of sources, which are believed to contain several spatially related fragments along the line of sight \citep{1990ApJ...359..355M}. 

A total of 193 OH thermal components are detected.  Both OH 1665 and 1667 emission were detected in 135 components, while OH 1665 and 1667 emissions were detected in 5 and 41 components alone, respectively.  The emission of one transition alone may arise from maser excitation instead of thermal emission. But this is difficult to distinguish since we don't have mapping observations.

The strength ratio between OH main lines ($R$=T$_b^{1667}$/T$_b^{1665}$) would lie in the range of  [1.0, 1.8] when  optical depth 
is considered under thermal excitation.  The strength ratio would deviate this range if non-thermal collision or radiation excitation dominates.It is called the OH main-line anomaly. 

\begin{figure}
\begin{center}  
  \includegraphics[width=0.48\textwidth]{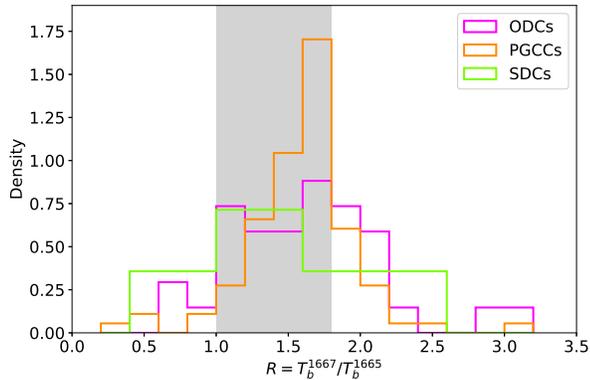}
  \caption{ Normalized density distribution of $R$ value for ODCs, PGCCs, and SDCs. The grey shade represents the $R$ range that can be explained by optical depth of OH under LTE. }
\label{fig:R_density}
\end{center}
\end{figure}

Statistical result of 135 OH components with detection of  two OH main lines is shown in Figure \ref{fig:R_density}. The percentage of OH main-line anomaly is  35.3\%,  20.9\%, and 33.3\% for ODCs, PGCCs and SDCs, respectively.  If the case with 1665 or 1667 detection alone is considered as OH main line anomaly, the percentage of OH main-line anomaly would be 60.7\%,  36.8\%, and 56.5\% for ODCs, PGCCs and SDCs, respectively. This may indicate that PGCCs are less affected by non-thermal effect or external irradiation compared to ODCs and SDCs. 


The satellite anomaly in which the satellite line strengths deviate from the LTE ratio is found to be common.  We follow \citet{1973ApJ...186..357T} and define the satellite anomaly in the following two types:

\begin{itemize}
\item[$\Pi$(a)]   Overpopulation of the F=1 level of $^2\Pi_{3/2}$ ground state. There are three observational cases for this type: 1) Enhanced 1612 MHz  emission and weaker even non-detected 1720 MHz emission when main lines exhibit as emission; 2) Non-detection of 1612 MHz line but there is obvious absorption  feature for both 1720 MHz and main-lines; 3) `Conjugate' behavior of the 1612 emission and  1720 absorption.  

\item[$\Pi$(b)]   Overpopulation of the F=2 level of $^2\Pi_{3/2}$ ground state. There are three observational cases for this type: 1) Enhanced 1720 MHz  emission and weaker even non-detected 1612 MHz emission when main lines exhibit as emission; 2) Emission or non-detection of 1720 MHz line but there is obvious absorption  feature for both 1612 MHz and main-lines; 3) `Conjugate' behavior of the 1612 absorption and 1720 emission. 

\end{itemize} 

The statistical result of OH satellite anomaly is presented in Table \ref{table:satellite-anomaly}. The results for ODC need to be checked further since  there is no 1720 MHz observations for ODCs.  

Collisional pumping could lead to enhancement of 1720 MHz line but require a relatively high kinetic temperature ($T_K\geq$ 25 K) to reach significant population reversion of ground F=1 and F=2 levels (e.g., Turner 1973).  The  kinetic temperatures of the molecular clouds are around 10 K, which means collisional pumping  may contribute little.  As seen in Fig. \ref{fig:oh_levels}, the  effect for satellite line anomaly is contributed  by far-infrared pumping and decay between the ground level state $^2\Pi_{3/2}$  and  higher energy levels of  OH  \citep[e.g., $^2\Pi_{5/2}$ and  $^2\Pi_{1/2}$;][]{1969ApJ...156..471L}.

As shown in Table \ref{table:satellite-anomaly},  Young Stellar Objects(YSO), Infrared sources (IR) or Far-infrared (FIR)  targets are found in the vicinity within radius of  2 arcmin toward  34 sources, almost 55\% of all sources with OH satellite anomaly. Sixteen of 18  sources with conjugate behavior contain  nearby YSOs, IR or FIR objects. This fraction is  much higher than the sub-type (1) and (2). This is reasonable since the overpopulation level would increase from sub-type (1) to (3).

Besides, these results imply that nearby sources emitting far-infrared photons are not necessary to lead to satellite anomaly. This is consistent with the conclusion by \citet{2000A&A...353.1065H} that a slightly more intense infrared radiation field than the general Galactic background could  lead to OH satellite-line anomaly.

\begin{table}
\caption{Statistical  number of  satellite anomaly types for ODC, PGCC and SDC. The number of sources with associated YSO, IR or FIR target in a radius of 2 arcmin are shown in brackets.\label{table:satellite-anomaly}}
\begin{tabular}{lllll}
\hline
Type  & Sub-type  &  ODC$^a$  & PGCC & SDC  \\
\hline
$\Pi$(a) & (1) &  2 (1)  &  8 (2) &   0 (0)  \\
$\Pi$(a) & (2) &  0 (0)  &  0 (0) &    0 (0)  \\
$\Pi$(a) & (3) &  0 (0)  &  4 (4) &   4 (4)  \\  
$\Pi$(b) & (1) &  1 (0)  & 30 (13) &  0 (0)   \\
$\Pi$(b) & (2) &  2 (2) &  1 (0)  &    0 (0)  \\
$\Pi$(b) & (3) &  0 (0) &  5 (3)  &    5 (5)  \\
\hline
\end{tabular}

{\raggedright $a$: Candidate, since there is no 1720 MHz data for ODCs. \par}
\end{table}

\subsection{OH Maser Emission}
\label{sec:maser-detection}

Three OH stellar maser with strong or multiple emission features  were clearly detected toward LDN649, SDC033.107-00.65 and SDC033.332-0.531. They are described in details as follows.

\begin{figure*}
\gridline{\fig{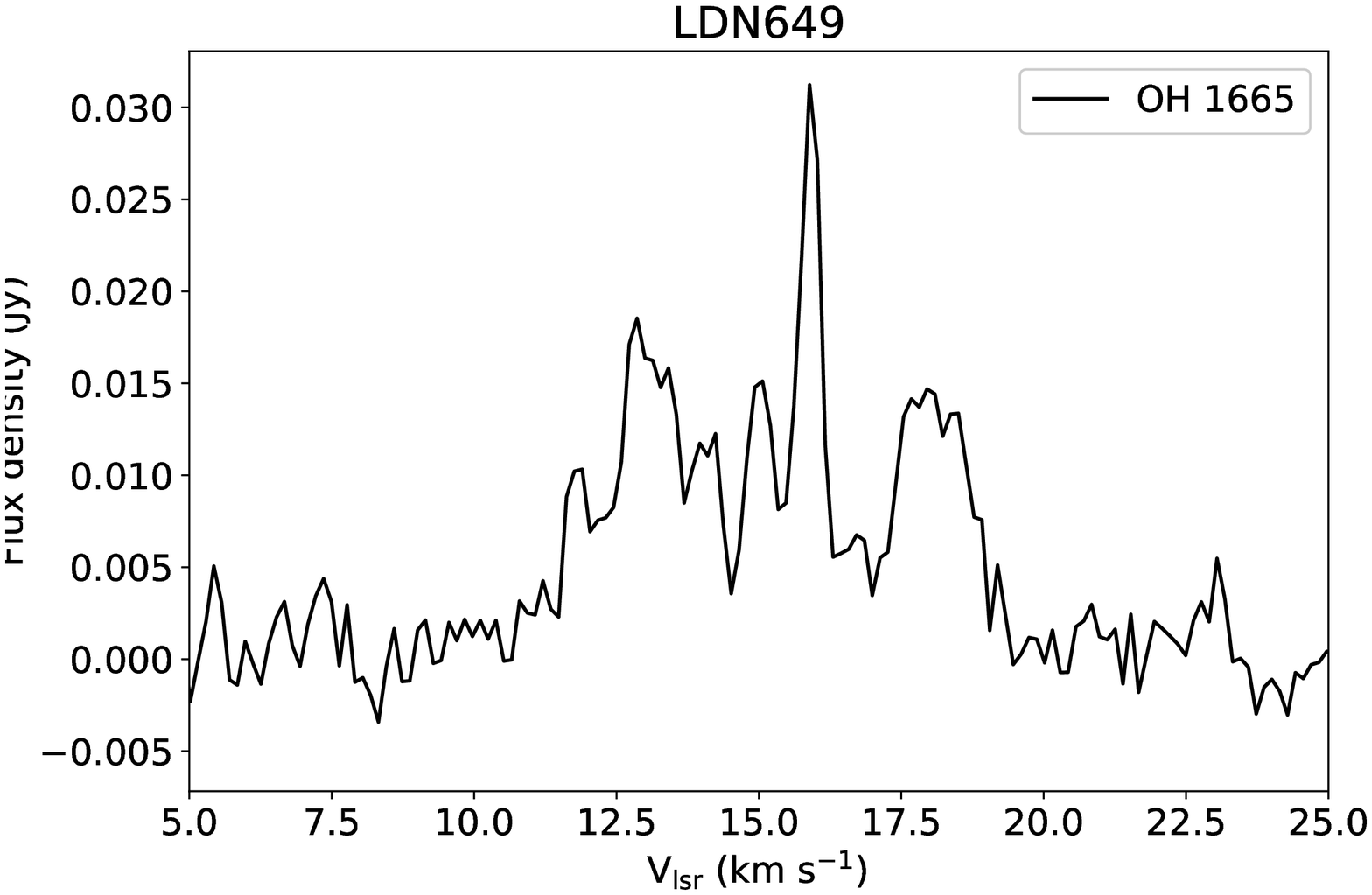}{0.33\textwidth}{(a)}
              \fig{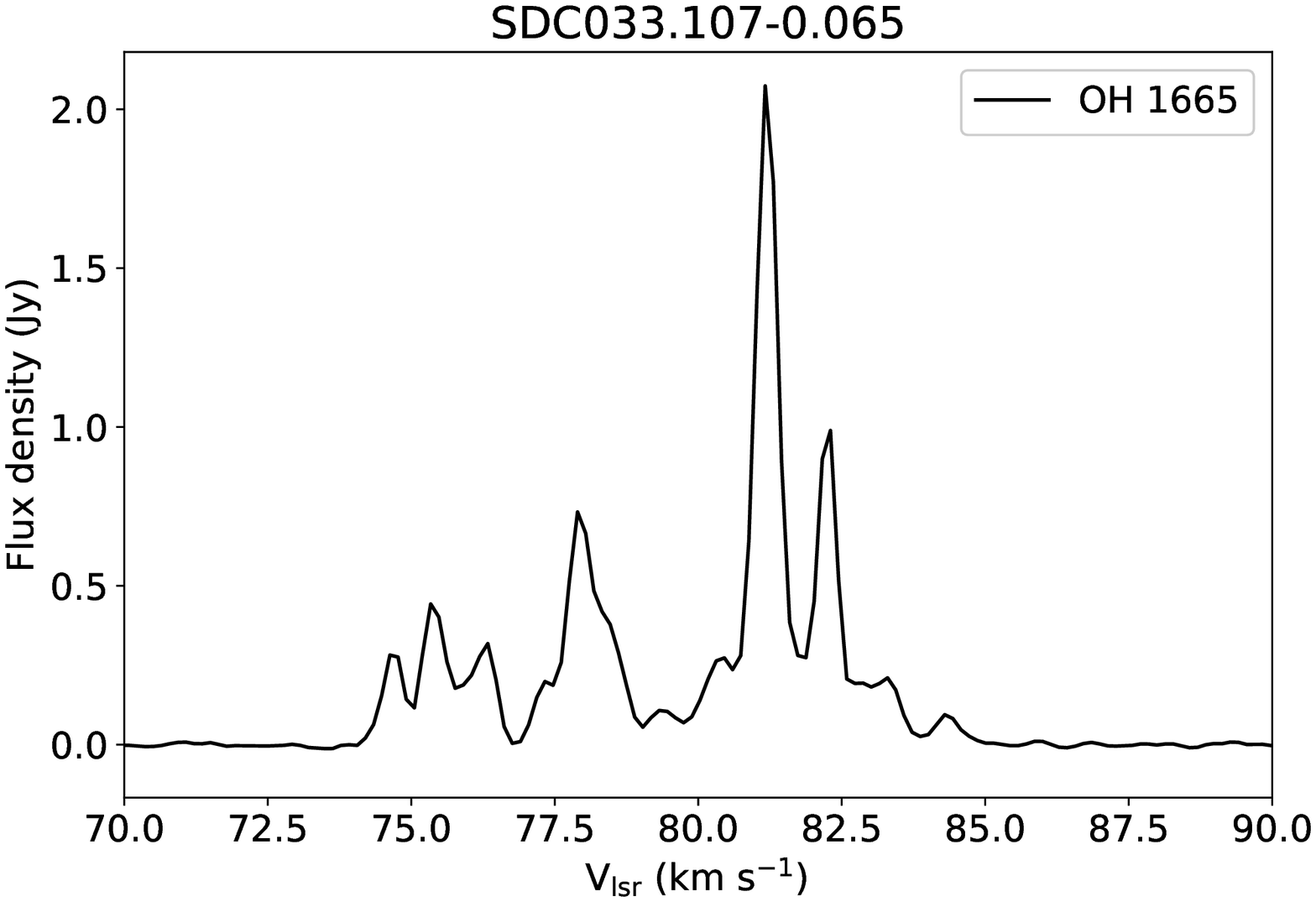}{0.33\textwidth}{(b)}
              \fig{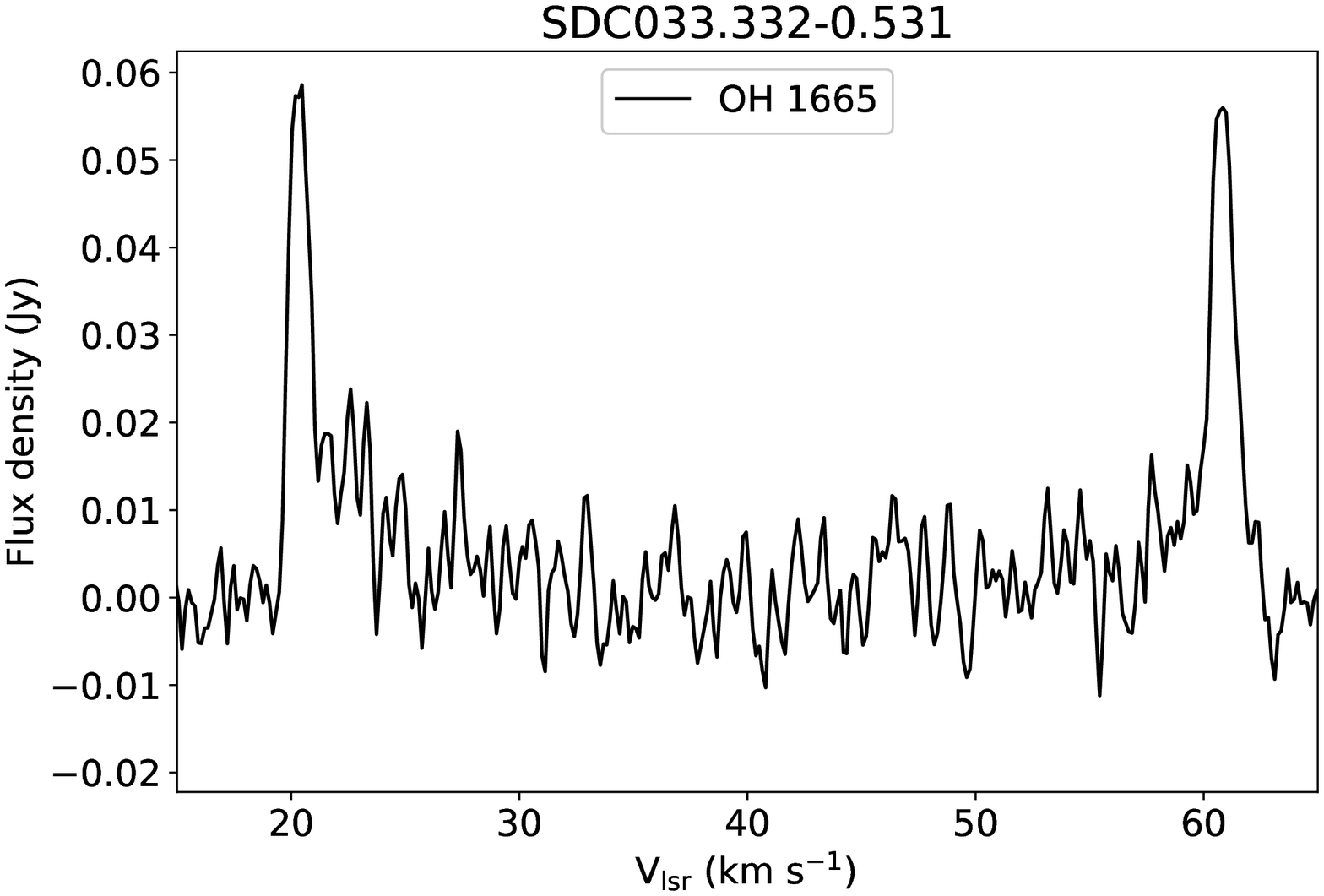}{0.33\textwidth}{(c)}}                       
\caption{Detected OH masers.  The flux density was converted from antenna temperature by dividing the Arecibo gain of 10 K/Jy.}
\label{fig:oh_maser_spectra}
\end{figure*}

$LDN649$: As shown in Figure \ref{fig:oh_maser_spectra}(a), this maser has a peak intensity of  $\sim$ 0.025 Jy  across a velocity range of 10-15 \kms\ in the OH 1665 spectrum.  A YSO locates 96 arcsec away from the source position. This maser may arise from the evolved stellar gas of this YSO. 

$SDC033.107-00.65$: As shown in Figure \ref{fig:oh_maser_spectra}(b), the maser emission appears in the velocity range of  75-85 \kms\ and has a peak of $\sim$ 2 Jy in the OH 1665 spectrum. It has typical feature of maser associated with high-mass star-forming regions and was reported in the previous OH survey of methanol sources \citep{2004A&A...414..235S}. This kind of masers are considered to arise from the the hot cores with high density ($\sim 10^7$ \cc) near ultracompact H II regions.

$SDC033.332-0.531$: In Figure \ref{fig:oh_maser_spectra}(c), this maser exhibits both blue (20 \kms) and red (60 \kms) peaks with magnitude of $\sim 0.06$ Jy, which belongs to typical maser features from an evolved stellar source. No known circumstellar maser was found with 2 arcmin radius of this source. It should be a new detection of circumstellar maser.

\section{Further Analysis}
\label{sec:analysis}

\subsection{Relation between W(OH) and A$\rm_V$}
\label{subsec:oh_cor}

The  $A\rm_V$ values of  ODCs were derived from SFD98, which is believed to trace the total proton along a sightline.
It contains two major gas components, \hi\ and \h2, thus $A\rm_V^{LOS}$ = $A\rm_V$(\hi)+$A\rm_V(H_2)$. 

To eliminate contribution of A$\rm_V$ value from the \hi\ emission along the sightline of the ODC, 
we derived  \hi\ intensity $I$(\hi)  by summing the \hi\ emission $T\rm_{mb}$  from -150 to 150 \kms\   $I$(\hi)= $\int_{-150}^{150} T\rm_{mb} d\upsilon$. $I$(\hi) was converted to HI column density ,  $N$(\hi)= 1.82$\times 10^{18} I$(\hi)/$fcor$  \cm2, in which $f\rm_{cor}$ represents the correction factor due to \hi\ optical depth $\tau$, $f\rm_{cor} = \int \tau d\upsilon/  \int (1-e^{-\tau}) d\upsilon $.  Previous absorption observations toward continuum sources show that $f\rm_{cor}$ value of 1-1.8  from \hi\ emission profile \citep[e.g.,][]{1982AJ.....87..278D}. Due to lack of prior information, we took $f\rm_{cor}=1$, which represents optically thin condition and  a lower limit of $N$(\hi) for ODCs.  The canonical conversion factor between N(\hi) and $A\rm_V$ , $A\rm_V$(\hi) = N(\hi)/1.87$\times 10^{21}$ mag was then adopted \citep{1975A&A....43..133S, 1978ApJ...224..132B}.  Uncertainties during this process are discussed  in Section \ref{subsec:uncertainty}.

Since there exist N(\h2) information for PGCCs and SDCs, the $A\rm_V(H_2)$ values of these clouds are directly derived from the N(H$_2$) through the transition $A\rm_V(H_2)$= 2N(H$_2$)/1.87$\times 10^{21}$ \cm2.

OH integrated intensity W(OH) was derived by summing the emission together along a line of sight. For SDCs with OH absorptions, we utilized the absolute intensity value of each component. 

\begin{figure}
\begin{center}  
  \includegraphics[width=0.48\textwidth]{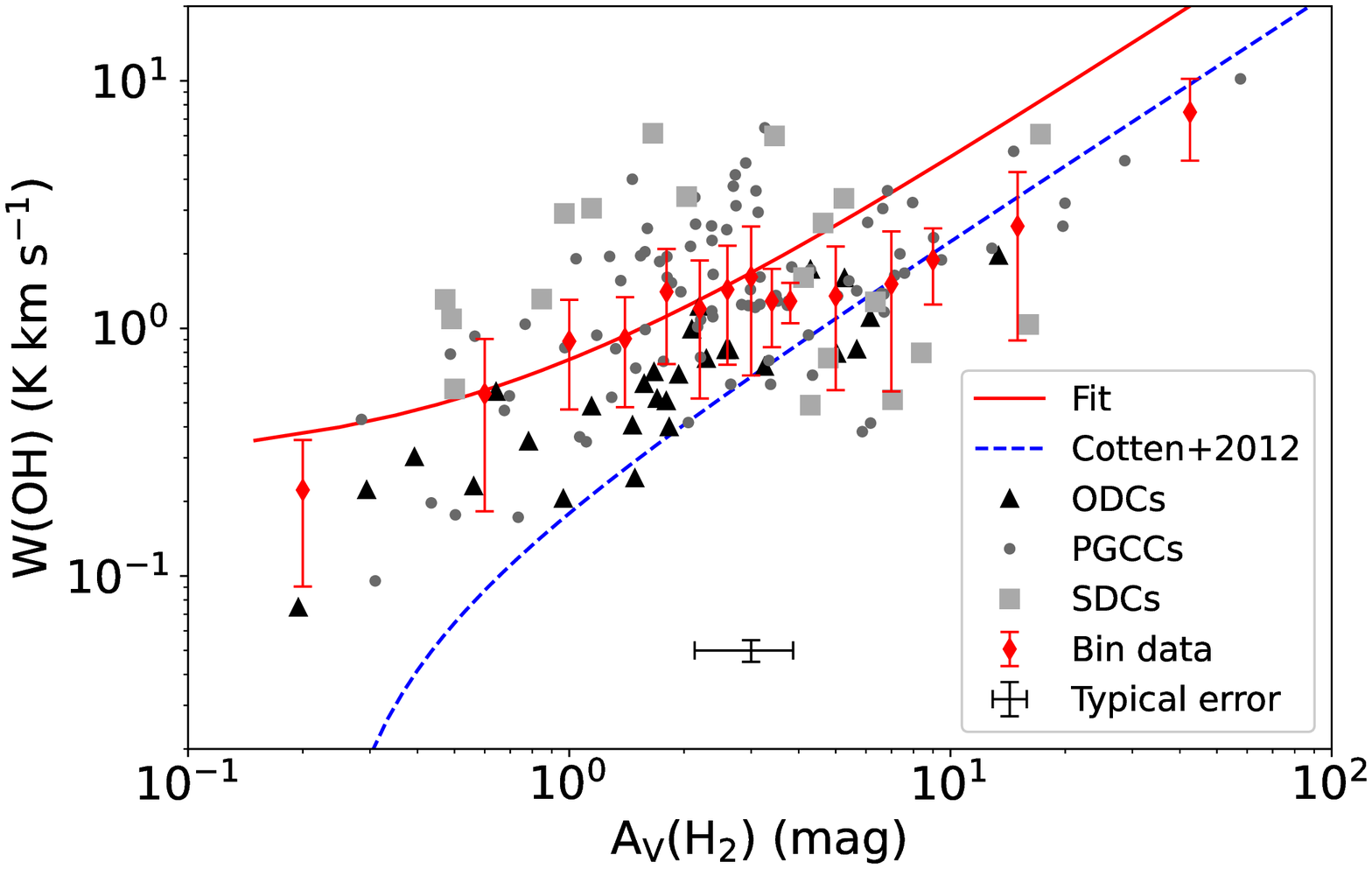}
  \caption{The relationship between W(OH) and A$\rm_V$. The data were binned with width of 0.4, 2 and 10 mag for $A\rm_V(H_2)$ range of [0, 4] mag,  [4, 10] mag and [10, 20] mag, respectively. The bin data at  $A\rm_V(H_2)$=42.5 mag is derived for the data with $A\rm_V(H_2)$ range of [20, 65] mag. Red solid line represents linear fit of  bin data with $A\rm_V \leq 3$ mag. The  correlation derived by Cotten et al.\ (2012) is shown with blue dashed line.  This correlation is valid for  $E(B-V)\rm (H_2)$ range of [0.08, 0.17] mag, which corresponds to  $A\rm_V(H_2)$ range of [0.25, 0.53] mag when $A\rm_V$/E(B-V)= 3.1 was adopted.}
\label{fig:woh_vs_ebv}
\end{center}
\end{figure}

In Figure \ref{fig:woh_vs_ebv}, we plot the the relation between OH intensity of 1667 MHz line, W(OH) and $A\rm_V^{H_2}$ along the sightline of three categories of clouds.  The W(OH) value of all OH diffuse components was summed together for each sightline.  OH maser components are excluded since they  significantly deviate  from collisional excitation. 

As shown in Figure \ref{fig:woh_vs_ebv}, W(OH)  correlates with  $A\rm_V(H_2)$  linearly following  W(OH)= (0.46 $\pm$ 0.06) $A\rm_V(H_2)$+(0.28$\pm 0.11$) when $A\rm_V(H_2)\leq 3$ mag.  It stays almost constant  at $A\rm_V(H_2)$ range of 3 to 7 mag and  increases linearly when  $A\rm_V(H_2)>7$ mag. Cotten et al.\ (2012) found an empirical relationship between  W(OH) and $E(B-V)\rm (H_2)$,  $W(\textrm{OH})  = (0.71\pm 0.08) E(B-V)\rm (H_2) - (0.05\pm 0.01)$ for the high latitude   translucent  cloud MBM40 with $E(B-V)\rm (H_2) \leq 0.17$ mag.  We plot this relationship in Figure \ref{fig:woh_vs_ebv} by assuming the conversion $A\rm_V$/E(B-V)= 3.1 \citep{1978ApJ...224..132B}.  The linear relationship  is consistent with our data only for $A\rm_V(H_2)>7$ mag and exists large discrepancy for the data with low $A\rm_V(H_2)$. 

\subsection{OH abundance in molecular clouds}
\label{subsec:oh_abundance}

The population of energy levels of OH main lines (1665 and 1667 MHz) is affected a little even there is strong satellite-line anomaly.  The intensity of 1667 MHz line is generally lager than that of the 1665 MHz line,  leading to a higher signal to noise ratio (SNR) for 1667 MHz line.  We calculate OH column density through the emission of  1667 MHz under the assumption of optically thin, which is confirmed by adsorptions toward quasars \citep[e.g.,][]{2018ApJS..235....1L}.  $N$(OH) can be derived by use of the formula \citep[e.g.,][]{1981A&A....98..271D}, 

\begin{equation}
N(\textrm{OH}) =  2.26 \times 10^{14} \frac{1}{f} \frac{T_\textrm{ex}}{T_\textrm{ex}- T_\textrm{bg}} \int T_\textrm{mb} d \upsilon\  \rm{cm^{-2}}, 
\label{eq:oh_column}
\end{equation} 
where $f$ is the beam filling factor. It is adopted as 1 since both the major and minor size of the observed sources are larger than the Arecibo beam of 2.7\arcmin$\times$3.1\arcmin\  at 1.6 GHz.  The OH absorption region may  be smaller than that of emission,  indicating a smaller filling factor and underestimation of N(OH) from absorption compared to that of emission \citep{2019ApJ...874...49E}. $T_\textrm{ex}$ and $T_\textrm{bg}$  are excitation temperature and background temperature of OH 1667 MHz transition, respectively; $T_\textrm{mb}$ is the main-beam temperature of 1667 MHz line. 

The value of $T_\textrm{ex}$ is critical for calculating N(\textrm{OH}). \citet{1968ApJ...151..919H} derived $T_\textrm{ex}$ by assuming local thermal equilibrium (LTE) between OH main lines, which implies $T^{ex}_{1667}=T^{ex}_{1665}$ and $\tau_{1667}= 1.8\tau_{1665}$.  The brightness temperature ratio ($R = T_b^{1667}/T_b^{1665}$) approaches 1 when $\tau_{1667}\gg 1$ and  1.8  when $\tau_{1667} \ll 1$.  Absorptions toward continuum sources convinced significant deviation of LTE even $R$ lies in the range of [1.0, 1.8] at small optical depth  \citep{1979ApJ...234..881C,1981A&A....98..271D,2018ApJS..235....1L}.  The adoption of LTE condition by assuming same excitation temperature of two main lines would lead to an overestimation of optical depth and then OH column density by more than one order of magnitude \citep[e.g.,][]{1977ApJ...216..308C}. 

Explicit measurement of $T\rm_{ex}$ were conducted through absorption/emission toward quasars \citep{1981A&A....98..271D,1996A&A...314..917L, 2018ApJS..235....1L} and grid mapping toward a continuum background, e.g., star-forming region \citep{2018ApJ...858...57E}.  

Due to lack of independent measurement of $T\rm_{ex}$ for our sources, we estimated the value of $T\rm_{ex}$/$(T\rm_{ex}$-$T\rm_{bg})$ through Monte Carlo simulation based on the fitted $T\rm_{ex}$ profile in \citet{2018ApJS..235....1L} in the section \ref{subsec:uncertainty}. Though there is no physical correlation between $T\rm_{ex}$ and $T\rm_{bg}$, the average value of $T\rm_{ex}$/$(T\rm_{ex}$-$T\rm_{bg})$ is consistent with the result by adopting  the value of $|T\rm_{ex}$-$T\rm_{bg}|$ $=$ 2.03 K. Since there is no direct measurement of continuum temperature with Arecibo telescope, the value of $T\rm_{bg}$ of each source was calculated  through the  equation  $T\rm_{bg}= 2.73 + (T_{bg408}-2.73)(\nu_{OH}/408 MHz)^{-2.7}$, in which $T_{bg408}$ is derived from the 408 MHz survey \citep{1982A&AS...47....1H}.  The value of $T\rm_{bg}$ ranges from 3.1 to 13.3 K,  resulting in  $T\rm_{ex}$/$(T\rm_{ex}$-$T\rm_{bg})$ range of [2.6, 7.7]. Uncertainties during the calculations are discussed in the section \ref{subsec:uncertainty}. 

Both radio and UV observations  suggest  that OH abundance $X$(OH) relative to H$_2$  \citep[e.g.,][]{1990ApJ...359..355M, 1996A&A...314..917L, 2010MNRAS.402.1991W, 2016ApJ...819...22X, 2018ApJ...862...49N}  ranges from $8\times 10^{-9}$ to 4$\times 10^{-6}$ and approaches  $10^{-7}$.   The X(OH)  range in our survey is from $5.7\times 10^{-8}$ to $4.8\times 10^{-6}$,  which is consistent with previous observations.  

\begin{figure}
\begin{center}  
\includegraphics[width=0.48\textwidth]{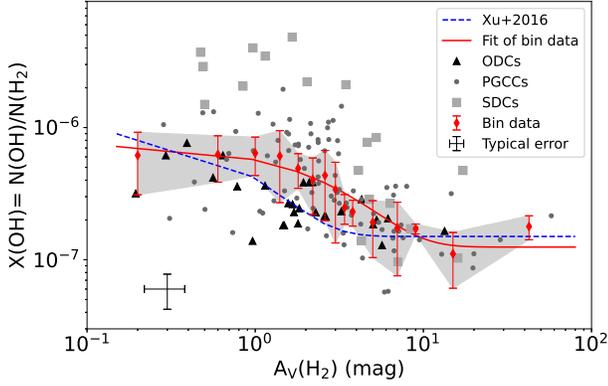}
\caption{The relationship between OH abundance and A$\rm_V$.   The range of bin data is the same as that in Fig. \ref{fig:woh_vs_ebv}.  The relationship found in \citet{2016ApJ...819...22X} is shown with blue dashed line. Fitting result of the bin data  is shown with red solid line.  As shown in Section \ref{subsec:uncertainty}, OH column density was calculated from the intensity of OH 1667 line by adopting excitation temperature derived from Monte Carlo simulations with one sigma uncertainty. }
\label{fig:X_oh_vs_ebv}
\end{center}
\end{figure}

Besides, the possible decreasing trend of  $X$(OH) in denser regions is suggested in previous observations. To investigate this in details, we plot  $X$(OH) as a function of  visual extinction $A\rm_V$ in Figure \ref{fig:X_oh_vs_ebv}.
$X$(OH)  stays almost constant before  decreasing $A\rm_V \sim 1.5$ mag.  It saturates again when $A\rm_V > 10$ mag.  \citet{2016ApJ...819...22X} found  an empirical relationship between X(OH)=[OH]/[H$_2$] and $A\rm_V$ across the boundary of Taurus molecular cloud, $X$(OH) = 1.5$\times 10^{-7}$ + 9.0$\times 10^{-7}\times$ exp(-A$\rm_V$/0.81), which is valid for  $A\rm_V$ range of 0.4 to 2.7 mag.  As a comparison, we plot this relationship in Figure \ref{fig:X_oh_vs_ebv}.  It is obvious that the trend of the bin data is consistent with the relationship from \citet{2016ApJ...819...22X}, but there exists significant discrepancy in the turn-off points.  Inspiring by this fact, we fitted  the bin data with same formula in \citet{2016ApJ...819...22X} and found the relationship follows, 

\begin{equation}
\frac{X(\textrm{OH})}{10^{-7}} = 1.3^{+0.4}_{-0.4} + 6.3^{+0.5}_{-0.5}\times \textrm{exp}(-\frac{A_\textrm{V}}{2.9^{+0.6}_{-0.6}}). 
\end{equation}

As for  chemistry evolution from the model, \citet{1986ApJS...62..109V} presented OH abundances ranging from $1.1\times 10^{-9}$ to $6.7\times 10^{-8}$  under 19 models with different density, kinetic temperature and visual extinction in diffuse and translucent clouds.  The visual extinction of all models  ranges from 0.64 to 1.01 mag expect one with $A\rm_V=2.12$ mag.  There is no obvious decreasing trend of  \xoh\ along with $A\rm_V$ from these modeling results. The reason may due to limited $A\rm_V$ range and varied density and kinetic temperature. \citet{1996A&A...311..690L} conducted photodissociation region (PDR) simulations of a static, plane-parallel, semi-infinite cloud with constant temperature of 30 K and density profile that represents self gravitating isothermal sphere. In the steady state of the simulations, OH abundance increases with $A\rm_V$ at $A\rm_V < 4$ mag and then decreases when $A\rm_V > 4$ mag. The peak is $\sim 8\times 10^{-8}$ corresponding to $A\rm_V = 4$ mag. 

\citet{2015MNRAS.450.4424B} found a maximum OH abundance of $\sim 6\times 10^{-7}$ near the H-to-\h2\ conversion region under low metallicities environment. In the \h2\ dominated region, OH abundance decrease linearly as increasing $\zeta/n$, in which $\zeta$ and $n$ are the cosmic ray ionization rate and total hydrogen density, respectively. As pointed out by the authors, this result is limited to the situations of low metallicities since  the dust-grain formation of the heavy molecules is neglected. \citet{2012ApJ...754..105H} described the detailed chemistry evolution in molecular clouds when polycyclic aromatic hydrocarbons (PAHs) were included in a steady-state PDR model. The OH abundance shows a peak of $\sim 3\times 10^{-6}$ at a $A\rm_V^{crit}$ value, which depends on the inputting  parameters in the model. 
$A\rm_V^{crit}$ $\sim 5$ mag when hydrogen density $n=10^4$ \cc, UV intensity  $\chi = 100$ and cosmic-ray ionization rate $\zeta= 2\times 10^{-16}$.

In a summary, the decreasing trend and saturation feature when $A\rm_V> 1.5$ mag in our data is consistent with previous model predictions. The increasing trend in predicted models at lower visual extinction (e.g., $A\rm_V< 1.5$ mag) is not clearly seen in this study. One possible explanation is that model always assume `fixed' physical parameters (e.g., temperature, density-extinction relation) which are unreasonable for analyzed clouds with totally different environments. 
Further quantitative comparison between the results and  PDR models are needed.  

\subsection{OH verseus CO}
\label{subsec:co_vs_oh}

\begin{figure}
\begin{center}  
  \includegraphics[width=0.48\textwidth]{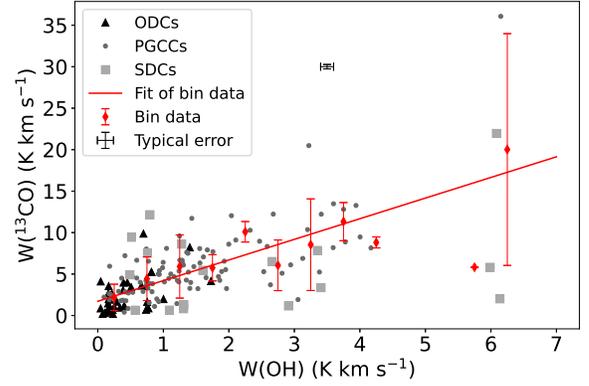}
  \caption{The relationship between intensity of  OH  and $^{13}$CO. The data were divided into bins with width of 0.5 K \kms.  The fitting result of bin data is shown with red solid line. } 
\label{fig:W13co-Woh}
\end{center}
\end{figure}

CO is the most common probe for tracing \h2\ in molecular regions.  CO spectral surveys \citep[e.g.,][]{2012ApJ...756...76W} toward dark clouds and dark clumps provide a direct statistics of star formation process. The J= 1-0 transition line of $^{13}$CO emission is optically thin even in cloud with extinction of  10 mag. It was detected  toward most of  the components with OH detection. We compare the \13co\ intensity W(\13co) with the W(OH) in Fig. \ref{fig:W13co-Woh}.  We can see that W(\13co)  correlates linearly with  W(OH) following the equation, W(\13co)= (2.49$\pm$ 0.40)W(OH) +(1.72 $\pm$ 1.26). 

The column density N(\13co) of \13co\ is calculated with the following equation, 
\begin{equation}
N(^{13}CO) = 2.42\times10^{14} \frac{T_x\tau^{13}\Delta V^{13}/0.937}{1.0-e^{-5.29/T_x}} cm^{-2}, 
\label{eq:13cocolden}
\end{equation}
where $T_{x}$ is excitation temperature of \13co. It was derived by assuming same excitation temperature of \co\ and \13co\ and optically thick of \co\  emission. $T_{x}= 5.532/ln(1+5.532/(T_b^{12}+0.819))$, in which $T_b^{12}$ is the brightness temperature of \co. The \13co\ optical depth $\tau^{13}=-log(1-T_b^{13}/5.29/J(T_x)-0.164)$, in which $T_b^{13}$ the brightness temperature of \13co\ and $J(T\rm_x)$ = 1/(exp(5.29/$T\rm_x$)-1).

\begin{figure}
\begin{center}  
  \includegraphics[width=0.48\textwidth]{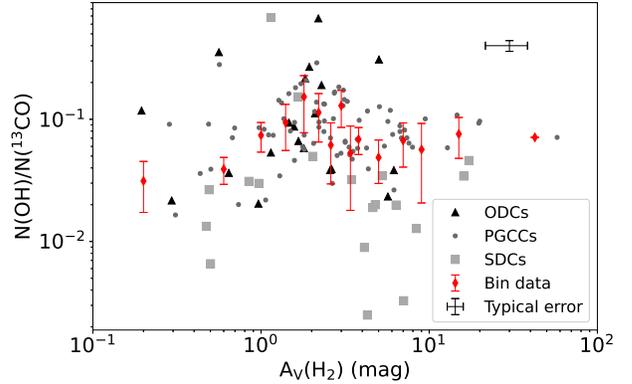}
  \caption{The relationship between  N(OH)/N($^{13}$CO) ratio and A$\rm_V$. The range of bin data is the same as that in Fig. \ref{fig:woh_vs_ebv}. }
\label{fig:oh213co_av}
\end{center}
\end{figure}

The N(OH)/N(\13co) ratio as a function of $A\rm_V(H_2)$ is shown in Figure \ref{fig:oh213co_av}. The N(OH)/N(\13co) ratio increases when $A\rm_V(H_2) <2$ mag.  It decreases until saturation at  $A\rm_V(H_2)=8$ mag. 

\section{Relationship between OH and  HINSA in Tracing Molecular Gas}
\label{sec:hinsa_analysis}

HINSA is widely used in tracing the cold \hi\ inside a molecular cloud. It was  found to be associated with 
molecular emissions in dark clouds \citep{2003ApJ...585..823L, 2010ApJ...724.1402K}. A total of 70 OH components
are associated with HINSA feature. An example of HINSA is shown around 5.8 \kms\ in Fig. \ref{fig:oh_spectra}(b). 

By adopting the radiation geometry in  \citet{2003ApJ...585..823L}, `expected' \hi\ profile without absorption can be expressed with the following equation, 

\begin{equation}
T_{\mathrm{H{\textsc i}}} = \frac{T_{\rm R} + (T_{\rm c}-T_{\rm k})(1-\tau_f)(1-e^{-\tau})}{1-p(1-e^{-\tau})}
\label{eq:hibackspec}
\end{equation}
where $T\rm_c$ represents the background continuum temperature contributed by the cosmic background and the Galactic continuum emission, $T_{\rm k}$ is the excitation temperature of the atomic hydrogen in the cold cloud, which is equal to the kinetic temperature, $\tau$ is the optical 
depth of the cold cloud. $\tau_f$ and $\tau_b$ are the optical depths of
warm \hi\ gas in front and behind the HINSA cloud. The total optical depth of warm \hi\ gas along the line of sight is 
$\tau_h=\tau_f+\tau_b$. $p$ is defined as the
fraction of background  \hi, $p=\tau_b/\tau_h$.  The value of $p$ depends on the spatial location of 
the  absorption cloud,  which is determined by  galactic longitude $l$,  galactic latitude $b$ and distance $d$ from the observer. Due to the difficulty in measuring cloud distance directly, we introduce  kinematic distance $d_\textrm{kin}$ which 
could be calculated through its velocity ($V_\textrm{cen}$) after applying  rotational curve.  Thus the value of $p$ is 
a function of $l$, $b$,  and $V_\textrm{cen}$,  with $p$= $p(l, b, V_\textrm{cen})$.

Once the cloud position is determined, three dimensional distribution of Galactic \hi\  is necessary to calculate $p$. \citet{2008A&A...487..951K}  derived  midplane \hi\ volume density distribution $n(R, z_0)$ and  \hi\ average thickness distribution $b_\textrm{R}$  as a function of Galactocentric radius $R$ based on  \hi\ surveys.  The descriptions of  $n(R, z_0)$ and $b_\textrm{z}(R)$ are shown in the following formula, 
 
 \begin{equation}
 n(R, z_0) \sim n_0 e^{-(R-R_\sun)/R_n} ,\  7\leq  R \leq 35\ kpc,
 \end{equation}
 where $n_0= 0.9$ cm$^{-3}$, $R_n$ = 3.15 kpc,  
 \begin{equation}
 b_z(R) \sim b_0e^{-(R-R_\sun)/R_0}  ,\  5\leq  R \leq 35\ kpc,
 \end{equation}
 where $b_0= 0.15$ kpc, $R_0$ = 9.8 kpc. 

By simply  adopting a gaussian distribution along $z$ axis, we constructed a 3-D  volume density distribution  $n(R,z)$ of \hi\ , which  can be described with

\begin{equation}
n(R, z) = n(R, z_0)e^{-z^2/b^2_z(R)}.
\end{equation}

With this distribution, we calculated the value of $p$ by considering 
integrated \hi\ content along the line-of-sight (LOS) of molecular cloud. 

\begin{figure}
\begin{center}  
  \includegraphics[width=0.48\textwidth]{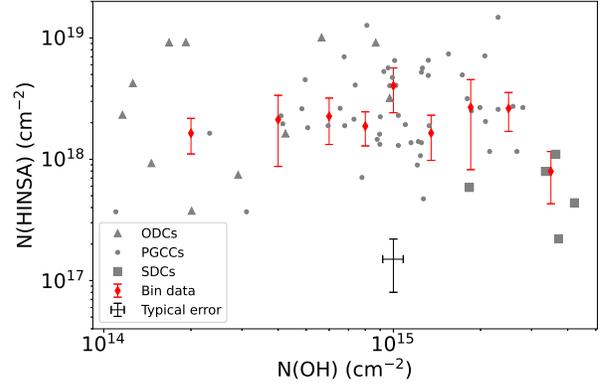}
\caption{ Relation between HINSA and OH column density.  The X values of bin data are [2, 4, 6, 8, 10, 13.5, 18.5, 25, 35]$\times 10^{14}$ \cm2. The limits of X bin data are [1, 3, 5, 7, 9, 11, 16, 21, 30, 40] $\times 10^{14}$ \cm2. } 
\label{fig:HINSA-OH}
\end{center}
\end{figure}

The value of  $p$  is 0.81 for LDN621.  As shown in Table \ref{table:HINSA_result} and Fig. \ref{fig:HINSA-OH}, the  HINSA column density  ranges from  2$\times 10^{17}$ to 2$\times 10^{19}$ \cc, with a median value of 2.3$\times 10^{18}$ \cc.   

As for the column density relationship between OH and HINSA, HINSA column density N(HINSA) stays almost constant when  $N\rm(OH)< 2.5\times 10^{15}$ \cm2.  

\section{Discussion}
\label{sec:discussion}

\subsection{Non-thermal Velocity Dispersions between OH, CO and HINSA }
\label{subsec:linewidth}

Non-thermal velocity broadening in molecular cloud is mainly caused by turbulence. Investigating the non-thermal 
line dispersion ($\sigma\rm_{NT}$) is beneficial to reveal kinematical behavior for different tracers.  We calculate $\sigma\rm_{NT}$ with the following equation, 

\begin{equation}
\sigma_\textrm{NT} =  \sqrt{\sigma_\textrm{obs}^2 - \sigma_\textrm{TH}^2}, 
\label{eq:nonth_width}
\end{equation}
in which $\sigma_\textrm{obs}=\Delta V\rm_{FWHM}/8ln(2)$ is the observed line dispersion. $ \sigma_\textrm{TH}= \sqrt{kT_k/m_H\mu}$ is the thermal velocity dispersion of the tracer. $k$, $T_k$, $m_H$ and $\mu$ are the Boltzmann constant, kinetic temperature, the mass of proton, and molecular weight of the tracer, respectively.  

\begin{figure*}
\gridline{\fig{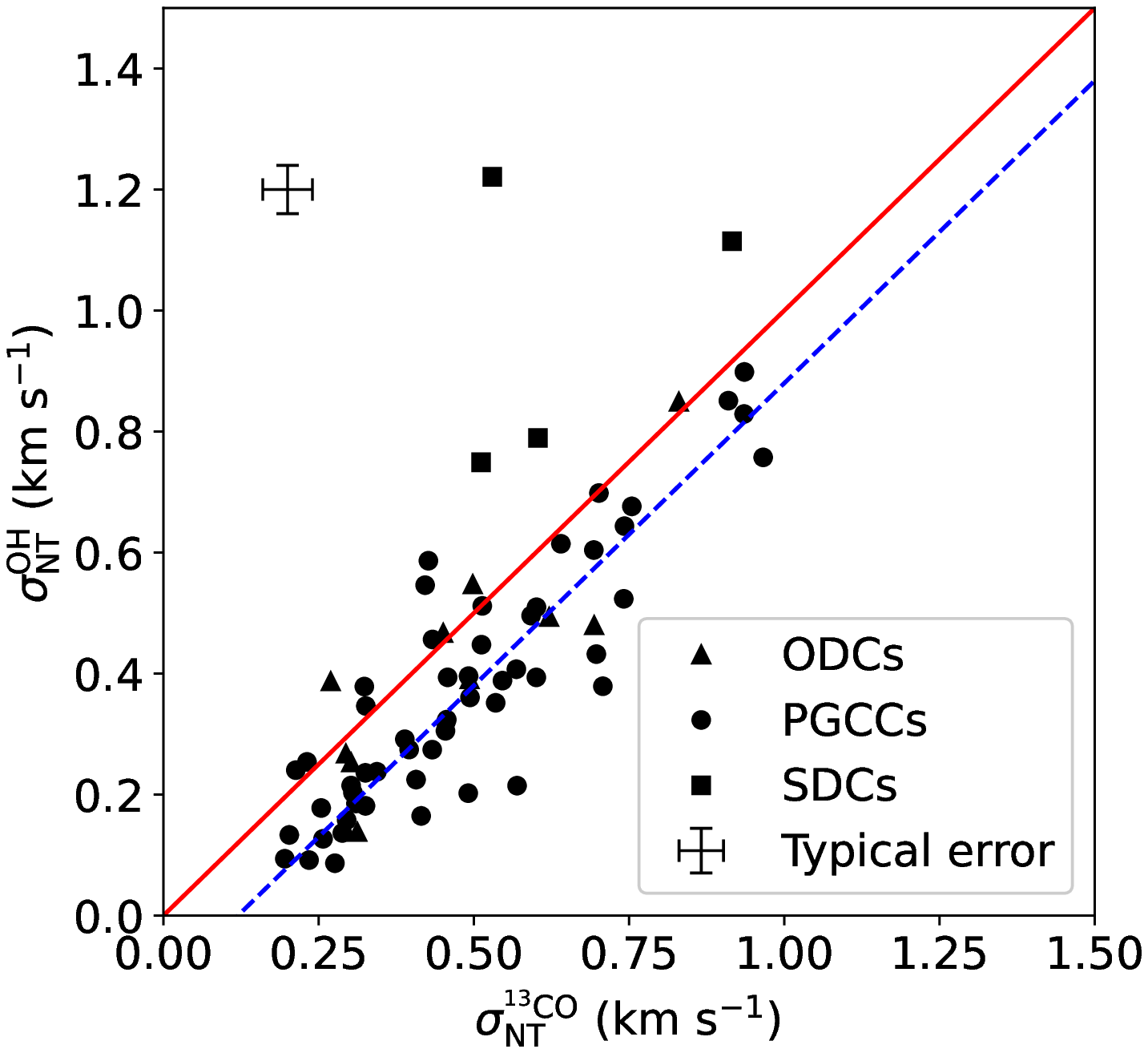}{0.48\textwidth}{(a)}
              \fig{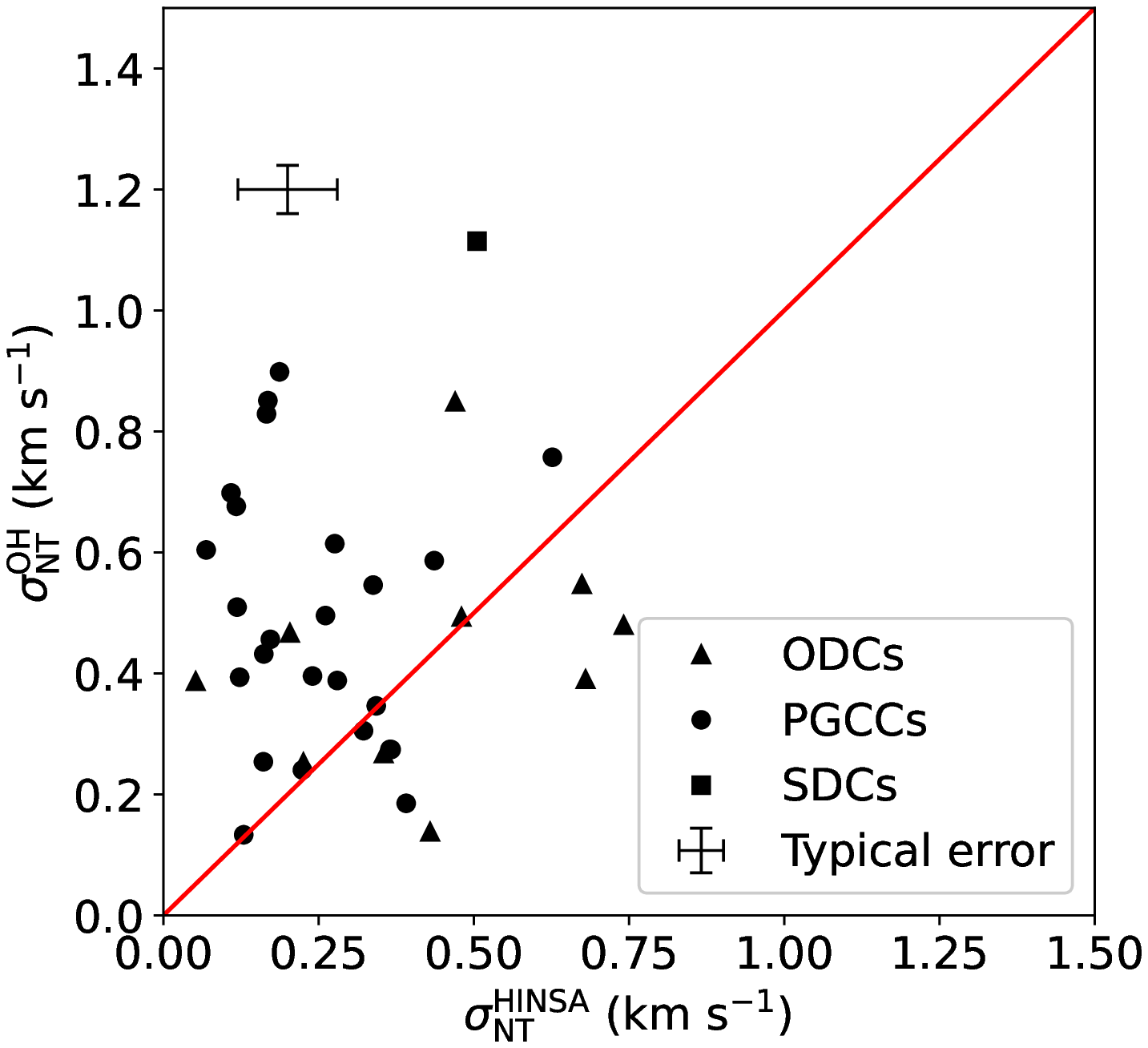}{0.48\textwidth}{(b)}}                       
\caption{ Comparison of nonthermal velocity dispersions of  different molecular tracers. (a):  OH versus \13co.  The red solid and blue dashed lines represent $\sigma\rm^{OH}_{NT}$= $\sigma\rm^{^{13}CO}_{NT}$ and  $\sigma\rm^{OH}_{NT}$= $\sigma\rm^{^{13}CO}_{NT}$-0.12 \kms, respectively. (b): OH versus HINSA. The red solid represents  $\sigma\rm^{OH}_{NT}$= $\sigma\rm^{HINSA}_{NT}$.}
\label{fig:nonth-wid}
\end{figure*}

By adopting  isothermal assumption of the ISM, we calculated the nonthermal velocity dispersions for \13co, OH and HINSA based on estimated value of $T\rm_k$ from CO observations.  Nonthermal velocities are generally less than $1.5$ \kms, consistent with previous CO observations (e.g., analysis of PGCCs in \citet{2012ApJ...756...76W}). Their relationships are shown in Fig. \ref{fig:nonth-wid}.  

As shown in Fig. \ref{fig:nonth-wid}(a),  $\sigma\rm^{OH}_{NT}$ is closely correlated with  $\sigma\rm^{^{13}CO}_{NT}$, implying similar volume occupation for OH and \13co\ in space.  The averaged value of $\sigma\rm^{OH}_{NT}$ is $\sim 0.12$ \kms\ smaller than that of $\sigma\rm^{^{13}CO}_{NT}$. \citet{2010MNRAS.407.2645B}  found broader nonthermal line width of OH compared to \co\ in cirrus clouds with high Galactic latitude. The discrepancy between our result and that in \citet{2010MNRAS.407.2645B}  may arise from limited sample in \citet{2010MNRAS.407.2645B}. 
 
No clear correlation is found for the nonthermal velocity dispersion between OH and HINSA. The  values of  $\sigma\rm^{OH}_{NT}$ are larger than that of $\sigma\rm^{HINISA}_{NT}$ is in  most clouds. Combined with the relationship between $N$(OH) and $N$(HINSA), OH and HINSA may occupy different volumes of space along sightline.  

In short, the averaged nonthermal dispersions follow  $\sigma\rm^{^{13}CO}_{NT}>$ $\sigma\rm^{OH}_{NT}$  $>\sigma\rm^{HINSA}_{NT}$.  The result of $\sigma\rm^{^{13}CO}_{NT}$  $>\sigma\rm^{HINSA}_{NT}$ is consistent with that found in \citet{2020RAA....20...77T}.

\subsection{  Cloud Age and Time dependent OH Abundance}
\hi\  Narrow Self-Absorption (HINISA) appears if there exists a warmer \hi\ background against the cold \hi\ gas along sightline. With an abundance of $\sim 10^{-2.8}$ \citep[e.g.,][]{2010ApJ...724.1402K} observed in HINSA survey of dark clouds, HINSA is another effective tracer of cold molecular gas in diffuse and dense molecular cloud.  

HINSA located in the central core of molecular clouds is balanced by formation through \h2 dissociation from cosmic ray and 
depletion of \hi-\h2\ transition. Thus it is a perfect tracer of the cloud `age', which indicates transition process between atomic and molecular gas. 

We adopted the same analysis method of cloud age in \citet{2005ApJ...622..938G}.  Time dependent \hi\ fraction is given by 
\begin{equation}
x_1(t) = 1- \frac{2k'n_0}{2k'n_0+\varsigma_{\rm H_2}}[1-exp(\frac{-t}{\tau_{\rm HI\rightarrow H_2}})]
\label{eq:hi-frac}
\end{equation}
in which $k'$ and $n_0$ are the formation rate coefficient of \h2\ and total proton density respectively. Grain size distribution would increase the formation rate coefficient by a factor of 3.4,  thus we adopt the  nominal value 
$k'=1.2\times 10^{-17}$ s$^{-1}$ \citep{2005ApJ...622..938G}).  $\varsigma_{\rm H_2}$ is the ionization rate of \h2\ by cosmic-ray, whose value ranges from $10^{-17}$ to 10$^{-15}$ s$^{-1}$ in dense and diffuse Galactic clouds \citep[e.g.,][]{1996ApJ...463..181F, 2000A&A...358L..79V, 2019ApJ...885..109B}, and even reaches $10^{-14}$ s$^{-1}$ in the Galactic center \citep{2016A&A...585A.105L}.  The time scale for atomic-molecular hydrogen conversion  $\tau_{\rm HI\rightarrow H_2}$ is described by 

\begin{equation}
\tau_{\rm HI\rightarrow H_2} = \frac{1}{2k'n_0+\varsigma_{\rm H_2}} .
\label{eq:hi-h2-timescale}
\end{equation}

The uncertainty of $\varsigma_{\rm H_2}$ has less effect for the analysis  since the densities of clouds are larger than that of 500 \cc\ and thus the term $2k'n_0\gg \varsigma_{\rm H_2}$ in Equation \ref{eq:hi-frac} and \ref{eq:hi-h2-timescale}. We adopted the value  $\varsigma_{\rm H_2}= 5.2\times 10^{-17}$ s$^{-1}$ from \citet{2000A&A...358L..79V} during calculation. 

\begin{figure}
\begin{center}  
\includegraphics[width=0.48\textwidth]{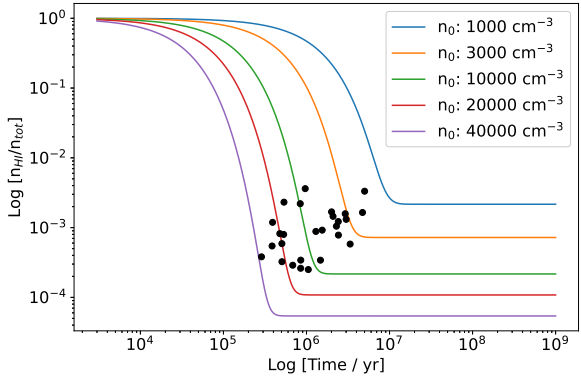}
\caption{Cloud age for PGCCs. Solid lines represent  \hi\ abundance as a function of cloud time in different proton densities,$1\times 10^3$, $3\times 10^3$, $1\times 10^4$, $2\times 10^4$, and $4\times 10^4$ \cc. }
\label{fig:cloud-age}
\end{center}
\end{figure}

Total proton volume density $n_0=n_\textrm{HI}+2n\rm{_{H_2}}$  is necessary to solve Equation \ref{eq:hi-frac}.  Thirty one PGCCs with both HINSA detection and  measured $n\rm_{H_2}$ values from Planck survey \citep{2016A&A...594A..28P} exist.  Since the size of \hi\ emission  is not available for each cloud, we adopt the reasonable assumption $n\rm_{HI}$ $\ll n\rm_{H_2}$ due to the fact that  $N$(HINSA)$\ll$ $N($\h2) \citep[e.g.,][]{2005ApJ...622..938G}.  Solutions of Equation \ref{eq:hi-frac} are available for 27 of 31 sources.  As shown in Figure \ref{fig:cloud-age}, the cloud ages of these sources range from $10^{5.5}$ to $10^{6.7}$ Myr. Most clouds are still undergoing \hi-\h2 transition and do not reach the steady state. 

\begin{figure}
\begin{center}  
\includegraphics[width=0.48\textwidth]{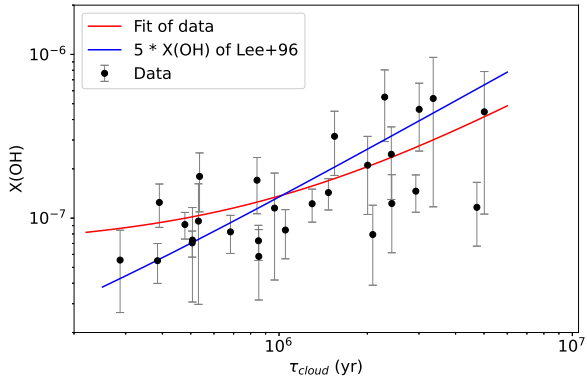}
\caption{Relationship between OH abundance X(OH) and cloud age $\tau_{cloud}$. The red line represents the fitting with linear function X(OH)= a*$\tau_{cloud}$+b, in which a= $(7.0 \pm 1.8) \times 10^{-14}$ and b=$(6.7\pm 3.7) \times10^{-8}$. The blue solid line represents the simulation results from \citet{1996A&A...311..690L}. A factor of 5 was multiplied to compare with our data.}
\label{fig:xoh-age}
\end{center}
\end{figure}

With the information of cloud age, it is available to derive the OH abundance as a function of time. This is important for comparison with time-dependent chemical model.  The time-dependent OH abundance is presented in Fig. \ref{fig:xoh-age}.  An obvious increasing trend was found for OH abundance during the time range of [0.29, 5.0] Myr.

We fit the results with a linear function. Best fitting parameters are described  with 
\begin{equation}
\frac{X(\textrm{OH})}{10^{-8}} = (7.0 \pm 1.8)\times \frac{\tau_{cloud}}{\textrm{Myr}}+ (6.7 \pm 3.7).
\label{eq:xoh-age_fit}
\end{equation}

Based on PDR simulations of a static, plane-parallel, semi-infinite cloud, \citet{1996A&A...311..690L} calculated time-dependent OH abundance and found a linear correlation between OH abundance and evolution time. This is valid for $A\rm_V$ range from 1.5 to 10 mag (maximum value of simulation, see Fig.\ 6a for details). In order to compare our data with this simulation, we fit a linear function of the OH abundances at age of $10^{4.5}$, $10^5$, $10^{5.5}$ and $10^{6}$ yr under the Model 1 (data are derived from Table 3 of  \citealt{1996A&A...311..690L}). As shown in Fig. \ref{fig:xoh-age}, our fitting is consistent with time-dependent simulation results except a factor of 5 in amplitude.

\subsection{OH as Tracer of Dark  Molecular Gas}
\label{sec:darkgas}

As a progenitor of CO formation, OH is considered as an effective tracer of `CO-dark' molecular gas(DMG).  This has 
been convinced by amount of  Galactic OH surveys \citep[e.g.,][]{2012AJ....143...97A,  2015AJ....149..123A, 2017ApJ...839....8T, 2019ApJ...883..158B}, Galactic OH absorptions toward Quasars \citep{2018ApJS..235....1L} or continuum sources \citep{2018A&A...618A.159R}, and envelope of molecular clouds \citep[e.g.,][]{2012AJ....144..163C, 2016ApJ...819...22X}.  

\begin{figure}
\begin{center}  
  \includegraphics[width=0.48\textwidth]{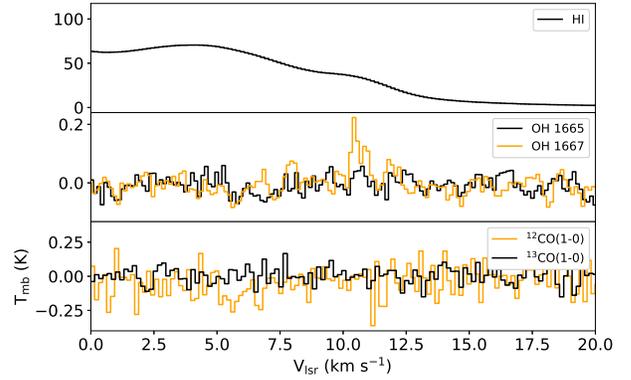}
  \caption{ Example of DMG detection toward LDN1525. The signal/noise ratio of OH 1667 emission at 10.2 \kms\  is $\sim$ 5.  }
\label{fig:LDN1525}
\end{center}
\end{figure}

In this survey,  we also detected 14 DMG components, which show OH emission  but lack corresponding CO emission.  As shown  in Fig. \ref{fig:LDN1525}, an example of DMG  happens for the OH 1667 emission with peak of 0.2 K at $\sim 10.2$ \kms toward LDN1425. These detections convince the validity of OH in tracing DMG. 

More cases were found toward CO emission without corresponding OH emission. This fact does not challenge the above conclusion and is understandable since OH excitation temperature is tightly close to the sbackground temperature. High sensitivity \citep[e.g, rms of $\sim$ 3 mK ;][]{2015AJ....149..123A} leads to a large possibility in detecting OH emission toward DMG.  In this work, OH sensitivity is only $\sim$ 42 mK for ODC and $\sim$ 88 mK for PGCCs and SDCs and gives a detection fraction of 7.3\%.
 
\subsection{Uncertainties}
\label{subsec:uncertainty}

During the above analysis, measurement uncertainties from the data are transferred into each variable by error propagation. 
Besides,  a series of uncertainties concerning about the calculation of N(OH) and N(\h2) exist. They are 
described as follows.
\begin{enumerate}
\item  The uncertainty during estimating continuum temperature $T\rm_C$. The continuum value we estimated in this paper is derived from the 408 MHz survey by adopting an power law index of -2.7.  This would certainly affect our calculation of  continuum temperature at 1.6 GHz for two reasons:  (a) The spatial resolution of 51\arcmin\ of 408 MHz survey is much larger than that of Arecibo beam of 3.1\arcmin\  during observations;  (b) The continuum contribution of H II region may dominate around the Galactic plane, especially for sources at the first quarter. As a rough check of the uncertainty,  the continuum level of  spectral bandpass ($T\rm_{cont}^{bp}$) of SDC033.382+0.204 (gl, gb = 33.382,0.204) is 36 K while it is 28 K for OFF plane source G049.76-07.25. It is difficult for us to tell the absolute continuum level toward the source  due to lack of  observation of a `clean' OFF position.  The value of $T\rm_{cont}^{bp}$ includes three main parts,  $T\rm_{cont}^{bp}$=    $T\rm_{sys}$ +$T\rm_{sky}$,  in which $T\rm_{sys}$ and $T\rm_{sky}$ represents system temperature of the telescope and sky temperature.  The value of $T\rm_{sys}$ depends on azimuth and zenith angle of the telescope. The minimum and maximum of $T\rm_{sys}$ values are $\sim$  22 K and $\sim 27$ K when a upper limit of 5 K is adopted.  Thus the continuum temperature ranges of SDC033.382+0.204 and G049.76-07.25 are [9, 14] K and [1, 6] K. Our  adopted $T\rm_C$ value of   SDC033.382+0.204 ($T\rm_C$=13.0 K) and  G049.76-07.25 ($T\rm_C$=4.7 K) locate in the value range from bandpass estimations. 

\item  The uncertainty of  $R=T_\textrm{ex}/(T_\textrm{ex}- T_\textrm{bg})$.  As described in section \ref{subsec:oh_abundance}, the $R$ value ranges from 2.6 to 7.7 under the adoption  of  $|T_\textrm{ex}- T_\textrm{bg} |=2.03$ K for all sources. To check the validity and uncertainty of this adoption,  we took Monte Carlo simulation and generated a sample of $T\rm_{ex}^{1667}$ that ranges from 3.0 to 25 K and satisfies the density probability formula,  $ \frac{1}{ \sqrt{2\pi}\sigma } \rm{exp}\left[-\frac{[ln(\textit{T}_{ex})-ln(3.4\ K)]^2}{2\sigma^2}\right] \lc$ in \citet{2018ApJS..235....1L}. The detection of  OH emission  implies  the equation  $T\rm_{ex} > $ $T\rm_{bg}$.  We assumed a lower limit of difference of 0.1 K,  $T\rm_{ex} -$ $T\rm_{bg}>$  0.1 K,  which is reasonable and may be underestimated  for rms of 0.044 K in obtained spectrum. As shown in Fig. \ref{fig:tex-tbg-ratio}, our adopted values are consistent  the averaged value from Monte Carlo simulation when $T\rm_{bg}< 13$ K.  The uncertainty of this adoption is large and may affect the calculation of N(OH).

 \begin{figure*}
\begin{center}  
\gridline{\fig{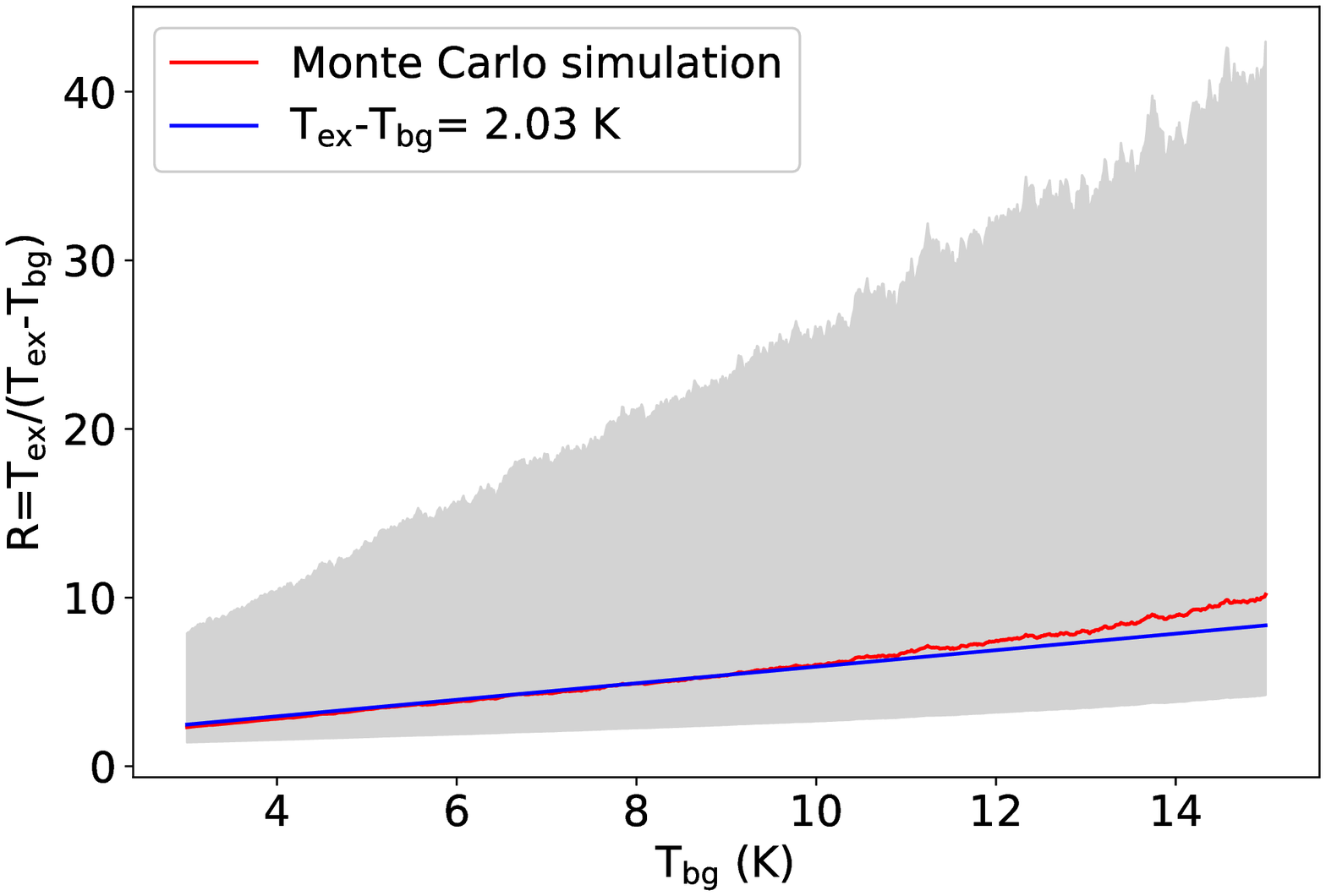}{0.49\textwidth}{(a)}
              \fig{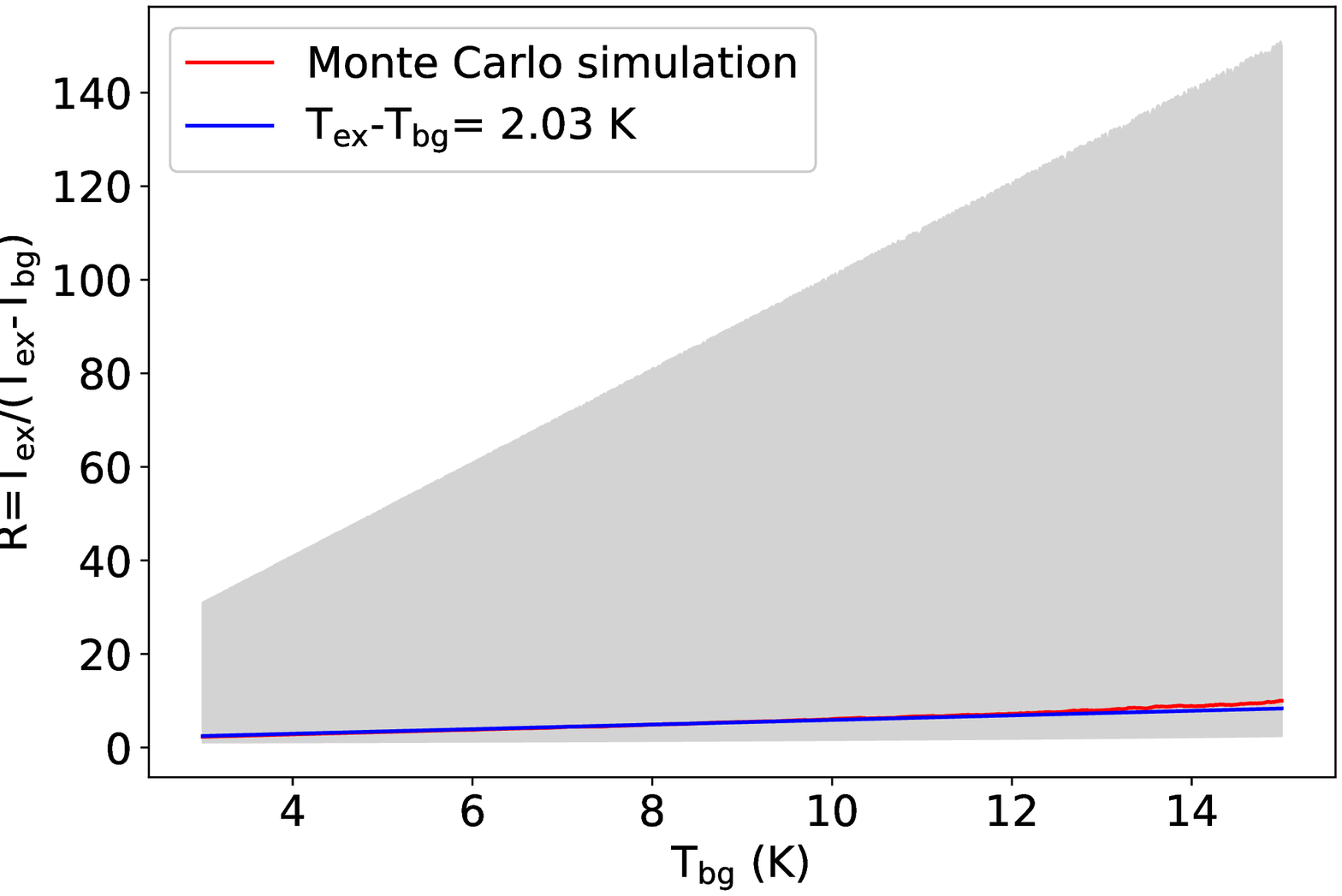}{0.49\textwidth}{(b)}}
  \caption{ The relationship between $R$ and $T\rm_{bg}$.  The results from Monte Carlo simulation and adoption of   $|T_\textrm{ex}- T_\textrm{bg} |=2.03$ K are shown with red and blue solid lines, respectively.  The  uncertainties with 66.7\% probability and 100\% probability from Monte Carlo simulation are shown in (a) and (b) with grey shade, respectively. }
\label{fig:tex-tbg-ratio}
\end{center}
\end{figure*}

\item  The uncertainty of conversion factor between proton column density  and A$\rm_V$. \citet{2018ApJ...862...49N} derived  N$\rm_H$/E(B-V) = $(9.4\pm 1.6)\times 10^{21}$ cm$^{-2}$  mag$^{-1}$ for pure 
\hi\ sightlines at high latitude ($|b|>5$ deg).  This value is 60\% higher than the canonical value  N$\rm_H$/E(B-V)= $5.8\times 10^{21}$ cm$^{-2}$  mag$^{-1}$ \citep{1978ApJ...224..132B}, which includes the contribution of both atomic and molecular gas.  For this consideration, we adopted the  canonical value to derive N(\hi) and  N(\h2)  for  SDC clouds  without direct N(\h2) measurements in the Galactic plane.  But we note that the derived X(OH) would be 38\% lower when the conversion factor in \citet{2018ApJ...862...49N} was adopted. 

\end{enumerate}

\section{Summary}
\label{sec:conclusions}

An OH 18 cm survey toward 33 ODCs, 98 PGCCs and 10 SDCs were made with Arecibo telescope.   With visual extinction ranging from 0.3 to 60 mag, these sources have different stages of cloud evolution. The survey toward these source reveals a comprehensive OH excitation and chemical evolution of molecular tracers for star formation. Our results are summarized as follows.

\begin{enumerate}
\item  A large fraction of OH components have 1667/1665 ratio  range deviating from the thermal ratio range of [1.0, 1.8], implying common non-thermal excitation in ODCs, PGCCs and SDCs. 

\item The infrared radiation from nearby YSOs, IR or FIR targets is not necessary to generate OH satellite anomaly except for  the satellite anomaly with `conjugate'  feature, which is always caused by infrared pumping.

\item OH intensity shows a linear correlation with visual extinction (Figure \ref{fig:woh_vs_ebv}), W(OH)= (0.46 $\pm$ 0.06)$A\rm_V$+(0.28$\pm 0.11$) when  $A\rm_V < 3$ mag. When $A\rm_V > 7$ mag,  W(OH)-$A\rm_V$ correlation  follows the linear relationship of \citet{2012AJ....144..163C}. 

\item  OH column density was calculated from the intensity of OH 1667 line by adopting excitation temperature derived from Monte Carlo simulations with one sigma uncertainty.  The  adoption of $|T_\textrm{ex}- T_\textrm{bg}| =2.03$ K is consistent with the average results of Monte Carlo simulations. 

\item The OH abundance was found to decrease with increasing visual extinction $A\rm_V$ (Figure \ref{fig:X_oh_vs_ebv}), 

\begin{equation}
\nonumber
\frac{X(\textrm{OH})}{10^{-7}} = 1.3^{+0.4}_{-0.4} + 6.3^{+0.5}_{-0.5}\times \textrm{exp}(-\frac{A_\textrm{V}}{2.9^{+0.6}_{-0.6}}),
\end{equation} 
consistent with PDR models. 

\item \13co\ intensity increases linearly with OH intensity following  W(\13co)= (2.49$\pm$ 0.40)W(OH) +(1.72 $\pm$ 1.26) (Figure \ref{fig:W13co-Woh}). 

\item  HINSA was detected toward 70 OH components. The column density of HINSA  locates in the  [$2\times 10^{17}$, $2\times 10^{19}$] \cm2 and has a median value of $2.3\times 10^{18}$ \cm2. HINSA column density stays almost constant when $N$(OH)$<2.5\times 10^{15}$ \cm2 (Figure \ref{fig:HINSA-OH}). 

\item Nonthermal velocity dispersion of OH is found to closely correlated with that of \13co (Figure \ref{fig:nonth-wid}a), indicating their being occupying similar spatial volumes. No obvious correlation is found between nonthermal velocity dispersion of OH and HINSA (Figure \ref{fig:nonth-wid}b). 

\item  Based on HINSA analysis of cloud age of 27 PGCCs with explicit density measurements, the relationship between OH abundance X(OH) and cloud age $\tau_{cloud}$ follows a linear function (Figure \ref{fig:xoh-age}),  
\begin{equation}
\frac{X(\textrm{OH})}{10^{-8}} = (7.0 \pm 1.8)\times \frac{\tau_{cloud}}{\textrm{Myr}}+ (6.7 \pm 3.7),
\end{equation}
which is consistent with time-dependent PDR simulations. 

\end{enumerate}

\section*{Acknowledgments}
We are grateful to the anonymous referee for his/her constructive suggestions, which have greatly improved this paper. 
This work is supported by the the National Natural Science Foundation of China (Grant No. 11988101, 11803051,and 11725313 ), National Key R\&D Program of China (2017YFA0402600),  CAS International Partnership Program (114A11KYSB20160008), CAS “Light of West China” Program.  S.-.L.Q. is supported by the Joint Research Fund in Astronomy (U1631237) under cooperative agreement between the National Natural Science Foundation of China (NSFC) and Chinese Academy of Sciences (CAS).

CO data were observed with the Delingha 13.7 m telescope of the Qinghai Station of Purple Mountain Observatory. We appreciate all the staff members of the Delingha observatory for their help during the observations.

\clearpage

\begin{longrotatetable}
\begin{deluxetable*}{lllllllllllllll}
\tablecaption{Gaussian fitting parameters of OH data.\label{table:ohfitresult}}
\tablewidth{700pt}
\tabletypesize{\tiny}
\tablehead{
\colhead{Srcid} & \colhead{Comp} &\colhead{Name} & \colhead{gl/gb} &  
\multicolumn{4}{c}{OH 1665 MHz} &  \multicolumn{4}{c}{OH 1667 MHz} & \colhead{T$\rm_k$} & \colhead{$N(^{13}$CO)} & \colhead{$N(\rm H_2)$} \\ 
\cline{5-7}
\cline{9-12}
\colhead{} & \colhead{} & \colhead{} & \colhead{}  & \colhead{$T\rm_{peak}$} & 
\colhead{$V\rm_{cen}$} & \colhead{$\Delta V$} &\colhead{} & \colhead{$T\rm_{peak}$} &
\colhead{$V\rm_{cen}$} & \colhead{$\Delta V$} & \colhead{$N$(OH)} & \colhead{} & \colhead{} & \colhead{}\\
\colhead{} & \colhead{}  & \colhead{} & \colhead{}  & \colhead{K} & 
\colhead{\kms} & \colhead{\kms} &\colhead{} & \colhead{K} &
\colhead{\kms} & \colhead{\kms} & \colhead{10$^{14}$ cm$^{-2}$} & \colhead{K} & \colhead{10$^{15}$ cm$^{-2}$} & \colhead{10$^{21}$ cm$^{-2}$} 
} 
\startdata
1 & 0 & CB124 & 35.12/11.37 & 0.08 (0.01) & 19.30 (0.11) & 1.16 (0.26) & & 0.06 (0.01) & 19.50 (0.39) & 2.59 (0.71) & 1.35 (0.44) & 6.96(0.14) & 1.67(0.16)  & 0.28(0.03) \\
1 & 1 & CB124 & 35.12/11.37 & \nodata & \nodata & \nodata &  & 0.06 (0.03) & 20.78 (0.13) & 0.76 (0.43) & 0.34 (0.25) & 8.56(0.10) & 6.16(0.24)  & 0.28(0.03) \\
2 & 0 & LDN621 & 35.35/2.34 & 0.16 (0.01) & 12.30 (0.11) & 2.19 (0.29) & & 0.12 (0.01) & 12.30 (0.15) & 2.18 (0.39) & 3.18 (0.69) & 3.75(0.31) & 2.41(0.25)  & 4.70(0.47) \\
2 & 1 & LDN621 & 35.35/2.34 & 0.13 (0.02) & 14.76 (0.11) & 1.37 (0.35) & & 0.28 (0.02) & 14.75 (0.05) & 1.30 (0.12) & 4.24 (0.49) & 4.55(0.29) & 0.00(0.00)  & 4.70(0.47) \\
2 & 2 & LDN621 & 35.35/2.34 & 0.09 (0.02) & 16.32 (0.13) & 0.92 (0.30) & & \nodata & \nodata & \nodata & \nodata &    \nodata & \nodata & 4.70(0.47) \\
2 & 3 & LDN621 & 35.35/2.34 & \nodata & \nodata & \nodata &  & 0.07 (0.02) & 27.16 (0.18) & 1.50 (0.42) & 1.27 (0.47) &  \nodata & \nodata & 4.70(0.47) \\
4 & 0 & CB191 & 37.56/-6.15 & \nodata & \nodata & \nodata &  & 0.09 (0.02) & 7.24 (0.12) & 0.91 (0.29) & 0.80 (0.32) &  \nodata & \nodata & 0.37(0.04) \\
4 & 1 & CB191 & 37.56/-6.15 & 0.07 (0.02) & 8.81 (0.15) & 1.32 (0.37) & & 0.21 (0.02) & 8.68 (0.05) & 0.94 (0.12) & 2.01 (0.33) & 7.31(0.10) & 5.74(0.40)  & 0.37(0.04) \\
6 & 0 & LDN638 & 39.33/3.55 & 0.07 (0.01) & 22.99 (0.20) & 2.54 (0.47) & & 0.05 (0.01) & 22.42 (0.32) & 3.53 (0.79) & 2.03 (0.58) &  \nodata & \nodata & 3.04(0.30) \\
6 & 1 & LDN638 & 39.33/3.55 & 0.07 (0.01) & 31.07 (0.18) & 2.01 (0.43) & & 0.09 (0.01) & 29.85 (0.23) & 5.15 (0.58) & 5.04 (0.73) &  \nodata & \nodata & 3.04(0.30) \\
7 & 0 & LDN640 & 40.14/-5.01 & \nodata & \nodata & \nodata &  & 0.13 (0.01) & 8.82 (0.08) & 1.67 (0.20) & 2.19 (0.35) & 6.19(0.07) & 0.62(0.16)  & 0.53(0.05) \\
9 & 0 & B140 & 40.68/-3.88 & 0.12 (0.02) & 6.08 (0.10) & 1.49 (0.25) & & 0.13 (0.01) & 6.56 (0.16) & 2.40 (0.40) & 3.60 (0.69) & 5.55(0.07) & 1.34(0.10)  & 1.81(0.18) \\
9 & 1 & B140 & 40.68/-3.88 & 0.11 (0.03) & 7.48 (0.05) & 0.38 (0.13) & & 0.11 (0.03) & 8.26 (0.07) & 0.69 (0.21) & 0.88 (0.34) &  \nodata & \nodata & 1.81(0.18) \\
9 & 2 & B140 & 40.68/-3.88 & \nodata & \nodata & \nodata &  & 0.07 (0.02) & 26.64 (0.13) & 0.87 (0.31) & 0.65 (0.30) &  \nodata & \nodata & 1.81(0.18) \\
9 & 3 & B140 & 40.68/-3.88 & \nodata & \nodata & \nodata &  & 0.06 (0.02) & 38.07 (0.15) & 0.89 (0.36) & 0.58 (0.31) &  \nodata & \nodata & 1.81(0.18) \\
9 & 4 & B140 & 40.68/-3.88 & \nodata & \nodata & \nodata &  & 0.10 (0.02) & 44.39 (0.10) & 1.12 (0.24) & 1.24 (0.35) &  \nodata & \nodata & 1.81(0.18) \\
10 & 0 & LDN645 & 42.66/-2.82 & \nodata & \nodata & \nodata &  & 0.22 (0.01) & 7.53 (0.06) & 1.70 (0.13) & 4.19 (0.44) & 7.87(0.10) & 1.94(0.15)  & 1.71(0.17) \\
11 & 0 & LDN649 & 43.03/-0.67 & \nodata & \nodata & \nodata &  & 0.11 (0.01) & 19.35 (0.12) & 1.87 (0.28) & 2.39 (0.47) &  \nodata & \nodata & 12.50(1.25) \\
11 & 1 & LDN649 & 43.03/-0.67 & 0.09 (0.02) & 58.39 (0.57) & 5.70 (1.41) & & 0.16 (0.01) & 57.38 (0.28) & 5.80 (0.57) & 10.51 (1.19) & 6.71(0.06) & 4.29(0.34)  & 12.50(1.25) \\
11 & 2 & LDN649 & 43.03/-0.67 & 0.06 (0.04) & 65.21 (0.42) & 1.46 (1.00) & & 0.11 (0.01) & 64.77 (0.45) & 6.48 (0.98) & 7.72 (1.31) & 5.69(0.09) & 1.32(0.16)  & 12.50(1.25) \\
12 & 0 & IREC58 & 43.22/8.33 & 0.11 (0.02) & 3.66 (0.09) & 0.93 (0.24) & & 0.15 (0.02) & 3.69 (0.10) & 0.96 (0.24) & 1.26 (0.35) & 10.50(0.45) & 2.96(0.36)  & 1.07(0.11) \\
12 & 1 & IREC58 & 43.22/8.33 & 0.15 (0.02) & 4.73 (0.06) & 0.61 (0.13) & & 0.26 (0.03) & 4.63 (0.05) & 0.65 (0.09) & 1.46 (0.26) & 9.03(0.27) & 2.92(0.21)  & 1.07(0.11) \\
12 & 2 & IREC58 & 43.22/8.33 & \nodata & \nodata & \nodata &  & 0.09 (0.01) & 10.58 (0.11) & 1.47 (0.27) & 1.16 (0.28) & 5.85(0.14) & 1.40(0.44)  & 1.07(0.11) \\
14 & 0 & LDN658 & 44.38/-2.33 & \nodata & \nodata & \nodata &  & 0.06 (0.02) & 7.55 (0.17) & 1.27 (0.40) & 0.85 (0.35) &  \nodata & \nodata & 1.39(0.14) \\
14 & 1 & LDN658 & 44.38/-2.33 & 0.07 (0.02) & 28.73 (0.13) & 1.02 (0.31) & & 0.13 (0.02) & 28.27 (0.07) & 1.15 (0.17) & 1.68 (0.33) & 6.71(0.06) & 5.09(0.15)  & 1.39(0.14) \\
16 & 0 & B335 & 44.93/-6.55 & 0.58 (0.03) & 8.03 (0.01) & 0.34 (0.02) & & 1.08 (0.03) & 8.05 (0.00) & 0.34 (0.01) & 3.26 (0.14) & 9.18(0.24) & 13.98(4.44)  & 5.31(1.69) \\
16 & 1 & B335 & 44.93/-6.55 & \nodata & \nodata & \nodata &  & 0.51 (0.03) & 11.29 (0.01) & 0.46 (0.03) & 2.08 (0.16) &  \nodata & \nodata & 5.31(1.69) \\
16 & 2 & B335 & 44.93/-6.55 & \nodata & \nodata & \nodata &  & 0.16 (0.02) & 24.73 (0.04) & 0.57 (0.10) & 0.80 (0.18) &  \nodata & \nodata & 5.31(1.69) \\
16 & 3 & B335 & 44.93/-6.55 & \nodata & \nodata & \nodata &  & 0.09 (0.02) & 38.40 (0.09) & 0.90 (0.22) & 0.70 (0.23) &  \nodata & \nodata & 5.31(1.69) \\
18 & 0 & B338 & 45.53/-8.05 & 0.07 (0.02) & 7.94 (0.12) & 0.78 (0.27) & & 0.08 (0.02) & 7.88 (0.10) & 0.86 (0.24) & 0.58 (0.21) & 6.10(0.10) & 0.49(0.10)  & 0.18(0.02) \\
19 & 0 & LDN672 & 45.79/-9.04 & 0.06 (0.02) & 20.38 (0.22) & 1.89 (0.52) & & 0.12 (0.02) & 20.65 (0.20) & 2.64 (0.47) & 2.62 (0.62) &  \nodata & \nodata & 0.73(0.07) \\
20 & 0 & LDN666 & 45.25/9.11 & 0.16 (0.06) & 5.27 (0.66) & 1.80 (0.82) & & 0.34 (0.02) & 5.65 (0.08) & 2.01 (0.11) & 5.65 (0.47) & 7.60(0.07) & 5.08(0.21)  & 1.96(0.20) \\
20 & 1 & LDN666 & 45.25/9.11 & 0.31 (0.15) & 6.32 (0.10) & 1.12 (0.21) & & 0.34 (0.04) & 6.40 (0.02) & 0.68 (0.08) & 1.92 (0.30) & 8.78(0.08) & 1.68(0.15)  & 1.96(0.20) \\
21 & 0 & LDN673 & 46.25/-1.30 & 0.34 (0.02) & 6.88 (0.02) & 0.74 (0.05) & & 0.57 (0.02) & 6.94 (0.02) & 1.15 (0.04) & 7.46 (0.34) & 12.56(0.11) & 19.46(0.45)  & 5.76(0.58) \\
21 & 1 & LDN673 & 46.25/-1.30 & 0.07 (0.01) & 24.43 (0.20) & 2.72 (0.47) & & 0.11 (0.01) & 24.80 (0.15) & 3.47 (0.37) & 4.35 (0.60) &  \nodata & \nodata & 5.76(0.58) \\
22 & 0 & LDN1484 & 165.72/-17.40 & 0.10 (0.01) & 7.10 (0.11) & 1.74 (0.28) & & 0.27 (0.02) & 6.88 (0.06) & 1.39 (0.13) & 2.49 (0.32) & 12.22(0.13) & 2.66(0.15)  & 1.37(0.14) \\
23 & 0 & LDN1485 & 166.02/-7.66 & \nodata & \nodata & \nodata &  & 0.14 (0.02) & -1.69 (0.09) & 1.16 (0.23) & 1.15 (0.28) & 9.28(0.20) & 1.50(0.25)  & 1.47(0.15) \\
23 & 1 & LDN1485 & 166.02/-7.66 & 0.15 (0.02) & -0.49 (0.06) & 1.14 (0.15) & & 0.23 (0.02) & -0.24 (0.05) & 0.90 (0.12) & 1.49 (0.24) & 10.52(0.12) & 1.73(0.21)  & 1.47(0.15) \\
23 & 2 & LDN1485 & 166.02/-7.66 & \nodata & \nodata & \nodata &  & 0.11 (0.02) & 2.00 (0.12) & 1.70 (0.30) & 1.29 (0.29) & 6.30(0.10) & 1.29(0.26)  & 1.47(0.15) \\
24 & 0 & LDN1486 & 166.55/-16.64 & 0.22 (0.02) & 6.25 (0.04) & 1.15 (0.10) & & 0.44 (0.02) & 6.24 (0.02) & 1.08 (0.04) & 3.16 (0.17) & 9.86(0.20) & 5.45(0.18)  & 1.68(0.17) \\
25 & 0 & LDN1489 & 167.96/-19.05 & 0.06 (0.02) & 5.81 (0.06) & 0.39 (0.14) & & 0.06 (0.01) & 5.72 (0.11) & 0.66 (0.26) & 0.26 (0.12) & 10.03(0.30) & 1.61(0.13)  & 0.90(0.09) \\
25 & 1 & LDN1489 & 167.96/-19.05 & 0.17 (0.01) & 6.65 (0.02) & 0.52 (0.05) & & 0.21 (0.01) & 6.62 (0.03) & 0.73 (0.08) & 0.99 (0.13) & 10.53(0.17) & 4.53(0.13)  & 0.90(0.09) \\
26 & 0 & MLB10 & 168.18/-16.28 & \nodata & \nodata & \nodata &  & 0.08 (0.01) & 4.16 (0.21) & 2.19 (0.55) & 1.19 (0.36) &  \nodata & \nodata & 4.93(0.49) \\
26 & 1 & MLB10 & 168.18/-16.28 & 0.55 (0.02) & 6.61 (0.02) & 1.25 (0.04) & & 1.10 (0.02) & 6.54 (0.01) & 1.20 (0.03) & 8.69 (0.25) & 12.47(0.12) & 15.84(0.37)  & 4.93(0.49) \\
27 & 0 & LDN1498 & 170.14/-19.11 & 0.40 (0.02) & 8.01 (0.02) & 0.79 (0.04) & & 0.88 (0.03) & 7.99 (0.01) & 0.71 (0.02) & 4.10 (0.18) & 9.34(0.11) & 6.19(0.23)  & 1.56(0.16) \\
28 & 0 & LM16 & 170.87/-15.87 & 0.40 (0.02) & 6.33 (0.02) & 1.02 (0.05) & & 0.72 (0.03) & 6.30 (0.02) & 1.06 (0.05) & 5.12 (0.34) & 9.44(0.10) & 13.34(0.77)  & 2.41(0.24) \\
29 & 0 & LM53 & 171.36/-10.72 & \nodata & \nodata & \nodata &  & 0.31 (0.02) & 5.80 (0.05) & 2.32 (0.13) & 5.01 (0.42) & 6.03(0.13) & 1.18(0.09)  & 2.05(0.20) \\
29 & 1 & LM53 & 171.36/-10.72 & 0.78 (0.03) & 5.94 (0.01) & 0.55 (0.02) & & 1.11 (0.03) & 5.95 (0.01) & 0.38 (0.01) & 2.91 (0.14) & 5.08(0.30) & 0.00(0.00)  & 2.05(0.20) \\
30 & 0 & LDN1508 & 171.62/-11.36 & 0.34 (0.01) & 5.53 (0.03) & 1.38 (0.07) & & 0.46 (0.01) & 5.48 (0.02) & 1.53 (0.05) & 4.89 (0.23) & 8.65(0.07) & 2.57(0.11)  & 2.14(0.21) \\
31 & 0 & MLB16 & 171.80/-15.42 & 0.40 (0.02) & 6.33 (0.02) & 1.02 (0.05) & & 0.72 (0.03) & 6.30 (0.02) & 1.06 (0.05) & 5.20 (0.35) & 9.44(0.10) & 13.34(0.77)  & 2.46(0.25) \\
32 & 0 & B222 & 172.79/-5.11 & 0.06 (0.02) & -6.06 (0.11) & 0.68 (0.26) & & 0.09 (0.02) & -5.93 (0.10) & 1.10 (0.25) & 0.72 (0.21) & 5.48(0.10) & 1.02(0.23)  & 0.60(0.06) \\
32 & 1 & B222 & 172.79/-5.11 & 0.22 (0.02) & 7.10 (0.03) & 0.49 (0.06) & & 0.44 (0.03) & 7.08 (0.01) & 0.45 (0.03) & 1.40 (0.14) & 13.03(0.11) & 4.78(0.10)  & 0.60(0.06) \\
32 & 2 & B222 & 172.79/-5.11 & \nodata & \nodata & \nodata &  & 0.18 (0.02) & 13.45 (0.05) & 1.21 (0.13) & 1.59 (0.23) &  \nodata & \nodata & 0.60(0.06) \\
33 & 0 & LDN1525 & 173.30/3.22 & \nodata & \nodata & \nodata &  & 0.06 (0.01) & -22.66 (0.26) & 3.66 (0.64) & 1.66 (0.38) &  \nodata & \nodata & 1.59(0.16) \\
33 & 1 & LDN1525 & 173.30/3.22 & 0.10 (0.02) & -5.96 (0.10) & 1.27 (0.23) & & 0.14 (0.02) & -5.91 (0.06) & 1.13 (0.15) & 1.16 (0.21) & 9.50(0.08) & 3.25(0.11)  & 1.59(0.16) \\
33 & 2 & LDN1525 & 173.30/3.22 & \nodata & \nodata & \nodata &  & 0.16 (0.02) & 10.51 (0.04) & 0.70 (0.11) & 0.82 (0.16) &  \nodata & \nodata & 1.59(0.16) \\
34 & 0 & G171.84-05.22 & 171.848/-5.230 & 2.56 (0.17) & 7.03 (0.01) & 0.38 (0.03) & & 4.31 (0.23) & 7.05 (0.01) & 0.37 (0.02) & 11.52 (0.98) & 11.82(0.35) & 9.86(1.06)  & 4.01(0.80) \\
35 & 0 & G173.45-05.43 & 173.452/-5.436 & 0.84 (0.27) & 6.75 (0.07) & 0.30 (0.12) & & 1.54 (0.11) & 6.73 (0.02) & 0.42 (0.03) & 4.64 (0.48) & 13.57(0.28) & 5.78(0.25)  & 1.40(0.57) \\
36 & 0 & G173.78-05.26 & 173.782/-5.267 & \nodata & \nodata & \nodata &  & 0.75 (0.12) & 7.51 (0.04) & 0.44 (0.07) & 2.34 (0.53) & 11.06(0.15) & 3.14(0.10)  & 1.04(0.55) \\
37 & 0 & G175.16-16.74 & 175.166/-16.744 & 2.19 (0.08) & 5.77 (0.02) & 0.84 (0.04) & & 3.94 (0.10) & 5.88 (0.01) & 0.80 (0.02) & 21.46 (0.84) & 7.95(0.90) & 15.55(13.91)  & 2.00(1.56) \\
38 & 0 & G175.58-16.60 & 175.583/-16.607 & 2.08 (0.10) & 5.75 (0.02) & 0.78 (0.04) & & 3.57 (0.10) & 5.74 (0.01) & 0.77 (0.03) & 18.54 (0.80) & 10.21(0.21) & 12.78(0.75)  & 2.93(1.23) \\
39 & 0 & G176.17-02.10 & 176.177/-2.108 & 2.06 (0.09) & -20.48 (0.02) & 0.85 (0.04) & & 3.33 (0.10) & -20.49 (0.01) & 0.74 (0.03) & 17.98 (0.81) & 10.26(0.19) & 6.21(0.31)  & 2.01(0.80) \\
40 & 0 & G176.35+01.92 & 176.352/1.921 & 0.66 (0.05) & -9.50 (0.08) & 1.97 (0.19) & & 1.10 (0.06) & -9.48 (0.04) & 1.66 (0.10) & 12.67 (1.03) & 8.11(0.16) & 8.87(0.39)  & 1.19(0.34) \\
41 & 0 & G176.37-02.05 & 176.374/-2.052 & 1.27 (0.08) & -20.37 (0.03) & 1.08 (0.08) & & 2.43 (0.07) & -20.21 (0.01) & 0.97 (0.03) & 17.28 (0.79) & 11.08(0.16) & 9.53(0.42)  & 1.50(0.77) \\
42 & 0 & G177.09+02.85 & 177.100/2.855 & 0.52 (0.05) & -10.12 (0.13) & 2.68 (0.31) & & 0.69 (0.05) & -10.27 (0.08) & 2.07 (0.19) & 9.93 (1.21) & 7.71(0.11) & 4.77(0.26)  & 1.73(0.45) \\
43 & 0 & G177.14-01.21 & 177.143/-1.213 & 1.14 (0.06) & -16.96 (0.04) & 1.78 (0.10) & & 2.12 (0.06) & -17.03 (0.02) & 1.65 (0.06) & 25.10 (1.11) & 12.43(0.13) & 24.81(0.57)  & 2.52(0.66) \\
44 & 0 & G177.86+01.04 & 177.869/1.044 & 0.60 (0.05) & -18.14 (0.10) & 2.28 (0.23) & & 0.92 (0.06) & -18.13 (0.05) & 1.58 (0.12) & 10.20 (1.00) & 8.45(0.44) & 10.09(0.39)  & 1.28(0.30) \\
45 & 0 & G178.28-00.61 & 178.286/-0.616 & 1.23 (0.11) & -0.79 (0.08) & 1.09 (0.16) & & 2.19 (0.07) & -0.79 (0.03) & 1.20 (0.07) & 18.60 (1.21) & 10.85(0.19) & 8.53(0.41)  & 2.72(0.64) \\
45 & 1 & G178.28-00.61 & 178.286/-0.616 & 0.71 (0.09) & 0.75 (0.15) & 1.45 (0.35) & & 1.39 (0.07) & 0.86 (0.04) & 1.24 (0.11) & 12.18 (1.24) & 11.88(0.18) & 8.23(0.38)  & 2.72(0.64) \\
46 & 0 & G178.48-06.76 & 178.484/-6.767 & 2.07 (0.07) & 7.23 (0.02) & 1.24 (0.05) & & 3.55 (0.10) & 7.18 (0.02) & 1.10 (0.04) & 28.03 (1.24) & 13.47(0.35) & 17.12(1.03)  & 2.55(1.18) \\
47 & 0 & G179.14-06.27 & 179.143/-6.279 & 2.04 (0.11) & 7.77 (0.02) & 0.63 (0.04) & & 2.95 (0.11) & 7.79 (0.01) & 0.62 (0.03) & 13.26 (0.72) & 8.54(0.30) & 0.00(0.00)  & 1.44(0.63) \\
48 & 0 & G179.29+04.20 & 179.297/4.200 & 1.11 (0.06) & 1.15 (0.04) & 1.66 (0.10) & & 1.86 (0.06) & 1.09 (0.03) & 1.80 (0.06) & 23.00 (1.06) & 9.51(0.21) & 13.27(0.97)  & 2.89(1.01) \\
49 & 0 & G181.16+04.33 & 181.164/4.331 & 0.32 (0.03) & 1.78 (0.08) & 1.71 (0.19) & & 0.56 (0.03) & 1.92 (0.06) & 2.07 (0.14) & 8.03 (0.70) & 10.65(0.20) & 15.90(0.73)  & 2.65(2.87) \\
50 & 0 & G181.71+04.16 & 181.714/4.163 & 0.51 (0.03) & 3.36 (0.05) & 1.70 (0.12) & & 0.89 (0.04) & 3.34 (0.04) & 2.02 (0.10) & 12.42 (0.81) & 10.56(0.18) & 10.74(0.51)  & 1.95(1.44) \\
50 & 1 & G181.71+04.16 & 181.714/4.163 & \nodata & \nodata & \nodata &  & 0.17 (0.05) & -6.18 (0.16) & 1.15 (0.38) & 1.38 (0.61) &  \nodata & \nodata & 1.95(1.44) \\
51 & 0 & G181.88+04.49 & 181.890/4.499 & 0.64 (0.03) & 5.30 (0.03) & 1.20 (0.07) & & 1.15 (0.05) & 5.09 (0.03) & 1.51 (0.07) & 11.93 (0.75) & 7.47(0.15) & 6.10(0.37)  & 1.61(1.31) \\
52 & 0 & G191.51-00.76 & 191.514/-0.765 & 2.04 (0.08) & -0.05 (0.02) & 1.13 (0.05) & & 3.49 (0.08) & -0.06 (0.01) & 1.01 (0.03) & 25.86 (0.98) & 13.67(0.26) & 18.47(0.81)  & 3.05(1.33) \\
52 & 1 & G191.51-00.76 & 191.514/-0.765 & 0.42 (0.12) & 3.23 (0.07) & 0.51 (0.15) & & 0.20 (0.03) & 4.68 (0.82) & 12.67 (2.01) & 18.34 (3.74) &  \nodata & \nodata & 3.05(1.33) \\
53 & 0 & G201.13+00.31 & 201.138/0.317 & 1.80 (0.07) & 5.19 (0.02) & 0.94 (0.04) & & 3.03 (0.07) & 5.17 (0.01) & 0.97 (0.03) & 20.79 (0.76) & 11.10(0.19) & 12.44(0.54)  & 2.56(0.26) \\
54 & 0 & G201.26+00.46 & 201.269/0.466 & 1.10 (0.06) & 5.31 (0.05) & 1.72 (0.12) & & 1.90 (0.06) & 5.30 (0.03) & 1.97 (0.07) & 26.69 (1.26) & 11.17(0.14) & 16.48(0.53)  & 1.37(1.56) \\
55 & 0 & G201.84+02.81 & 201.841/2.817 & 0.73 (0.06) & 5.90 (0.04) & 0.93 (0.09) & & 1.22 (0.05) & 5.89 (0.02) & 1.12 (0.06) & 9.59 (0.65) & 13.63(0.15) & 11.58(0.39)  & 0.97(0.52) \\
55 & 1 & G201.84+02.81 & 201.841/2.817 & \nodata & \nodata & \nodata &  & 0.30 (0.05) & 8.78 (0.11) & 1.41 (0.27) & 2.91 (0.73) &  \nodata & \nodata & 0.97(0.52) \\
56 & 0 & G158.40-21.86 & 158.401/-21.864 & 0.92 (0.05) & 4.33 (0.03) & 1.08 (0.06) & & 1.78 (0.05) & 4.25 (0.01) & 0.88 (0.03) & 10.38 (0.45) & 11.98(0.16) & 16.29(0.57)  & 7.08(4.11) \\
57 & 0 & G158.86-21.60 & 158.862/-21.603 & 0.90 (0.05) & 4.89 (0.03) & 0.97 (0.06) & & 1.36 (0.05) & 4.92 (0.02) & 1.07 (0.04) & 9.71 (0.52) & 14.84(0.18) & 16.11(0.82)  & 5.06(3.48) \\
59 & 0 & G159.78-24.80 & 159.785/-24.809 & 0.28 (0.07) & 2.71 (0.05) & 0.42 (0.11) & & 0.23 (0.07) & 3.46 (0.06) & 0.39 (0.15) & 0.60 (0.30) & 10.42(0.29) & 3.61(0.50)  & 0.29(0.08) \\
60 & 0 & G168.00-15.69 & 168.003/-15.695 & 1.40 (0.06) & 7.75 (0.01) & 0.59 (0.03) & & 2.43 (0.06) & 7.77 (0.01) & 0.62 (0.02) & 10.07 (0.38) & 13.80(0.32) & 15.56(0.62)  & 2.96(1.11) \\
61 & 0 & G168.13-16.39 & 168.135/-16.393 & 2.13 (0.07) & 6.43 (0.01) & 0.64 (0.02) & & 4.01 (0.08) & 6.44 (0.01) & 0.60 (0.01) & 15.98 (0.48) & 11.70(0.28) & 17.20(1.02)  & 18.44(15.94) \\
62 & 0 & G168.72-15.48 & 168.728/-15.482 & \nodata & \nodata & \nodata &  & 0.48 (0.04) & 6.07 (0.05) & 1.12 (0.14) & 3.63 (0.56) &  \nodata & \nodata & 6.21(1.83) \\
62 & 1 & G168.72-15.48 & 168.728/-15.482 & 2.51 (0.06) & 7.30 (0.01) & 0.57 (0.02) & & 4.03 (0.06) & 7.33 (0.00) & 0.57 (0.01) & 15.53 (0.38) & 12.93(0.25) & 21.76(2.51)  & 6.21(1.83) \\
63 & 0 & G169.82-19.39 & 169.827/-19.392 & 0.95 (0.06) & 8.08 (0.02) & 0.62 (0.04) & & 2.41 (0.06) & 8.13 (0.01) & 0.48 (0.01) & 7.59 (0.31) & 13.45(0.17) & 11.08(0.52)  & 3.49(2.03) \\
64 & 0 & G169.84-07.61 & 169.849/-7.613 & 0.86 (0.07) & 6.43 (0.02) & 0.44 (0.04) & & 1.50 (0.07) & 6.38 (0.01) & 0.47 (0.03) & 5.06 (0.38) & 10.92(0.11) & 5.87(0.19)  & 2.07(0.45) \\
65 & 0 & G170.77-08.51 & 170.771/-8.518 & 0.92 (0.07) & 6.83 (0.02) & 0.41 (0.04) & & 1.17 (0.07) & 6.79 (0.02) & 0.59 (0.04) & 4.83 (0.44) & 12.16(0.28) & 8.11(0.49)  & 1.65(0.40) \\
66 & 0 & G170.88-10.92 & 170.881/-10.921 & 0.86 (0.07) & 5.98 (0.02) & 0.49 (0.04) & & 1.67 (0.08) & 5.99 (0.01) & 0.36 (0.02) & 4.18 (0.30) & 8.74(0.16) & 10.66(1.42)  & 4.06(1.34) \\
67 & 0 & G171.34-10.67 & 171.343/-10.674 & 1.24 (0.06) & 6.26 (0.02) & 0.68 (0.04) & & 1.87 (0.06) & 6.29 (0.01) & 0.69 (0.03) & 8.96 (0.44) & 7.87(0.43) & 0.00(0.00)  & 6.16(1.46) \\
68 & 0 & G173.07-16.52 & 173.079/-16.529 & 1.35 (0.06) & 6.69 (0.01) & 0.56 (0.03) & & 2.34 (0.07) & 6.71 (0.01) & 0.51 (0.02) & 8.10 (0.34) & 14.41(0.29) & 14.00(0.84)  & 3.51(2.23) \\
69 & 0 & G173.07-17.89 & 173.079/-17.896 & 0.51 (0.04) & 3.91 (0.05) & 1.05 (0.11) & & 0.75 (0.06) & 4.06 (0.03) & 0.74 (0.06) & 3.77 (0.42) & 12.89(0.18) & 12.60(0.46)  & 2.48(0.55) \\
70 & 0 & G173.12-13.32 & 173.122/-13.325 & 1.23 (0.06) & 5.41 (0.02) & 0.62 (0.04) & & 2.01 (0.06) & 5.37 (0.01) & 0.65 (0.02) & 9.01 (0.41) & 8.66(1.20) & 0.00(0.00)  & 1.83(0.86) \\
70 & 1 & G173.12-13.32 & 173.122/-13.325 & 0.50 (0.05) & 6.50 (0.06) & 0.99 (0.15) & & \nodata & \nodata & \nodata & \nodata &   7.21(0.45) & 4.29(1.26)  & 1.83(0.86) \\
71 & 0 & G173.36-16.27 & 173.364/-16.277 & 1.85 (0.06) & 6.35 (0.01) & 0.54 (0.02) & & 3.17 (0.07) & 6.35 (0.01) & 0.59 (0.01) & 12.62 (0.40) & 12.22(0.40) & 17.90(2.72)  & 6.89(1.26) \\
72 & 0 & G173.69-15.55 & 173.694/-15.559 & 1.06 (0.05) & 6.30 (0.02) & 0.67 (0.04) & & 1.72 (0.06) & 6.28 (0.01) & 0.63 (0.02) & 7.31 (0.38) & 11.10(0.31) & 9.21(0.81)  & 6.26(2.85) \\
73 & 0 & G173.91-16.25 & 173.913/-16.257 & 0.83 (0.05) & 6.09 (0.03) & 0.86 (0.06) & & 1.87 (0.06) & 6.17 (0.01) & 0.65 (0.02) & 8.18 (0.41) & 15.35(0.34) & 14.96(0.90)  & 3.29(2.13) \\
74 & 0 & G173.95-13.74 & 173.957/-13.747 & 2.17 (0.05) & 6.31 (0.01) & 1.08 (0.03) & & 2.66 (0.05) & 6.25 (0.01) & 1.27 (0.03) & 23.03 (0.65) & 12.97(0.28) & 32.77(2.57)  & 6.39(2.50) \\
75 & 0 & G174.06-15.81 & 174.067/-15.811 & 1.61 (0.05) & 6.23 (0.01) & 0.86 (0.03) & & 2.52 (0.05) & 6.26 (0.01) & 0.86 (0.02) & 14.70 (0.48) & 13.40(0.17) & 25.98(1.00)  & 8.44(2.14) \\
76 & 0 & G174.39-13.43 & 174.397/-13.440 & 1.88 (0.05) & 5.81 (0.01) & 1.19 (0.03) & & 3.19 (0.06) & 5.75 (0.01) & 0.94 (0.02) & 20.71 (0.55) & 11.79(0.27) & 21.47(0.91)  & 18.68(17.11) \\
77 & 0 & G174.44-15.75 & 174.441/-15.753 & 0.80 (0.06) & 5.27 (0.02) & 0.58 (0.05) & & 1.04 (0.05) & 5.23 (0.02) & 0.81 (0.05) & 5.68 (0.47) & 7.98(0.29) & 2.94(0.56)  & 1.69(1.46) \\
77 & 1 & G174.44-15.75 & 174.441/-15.753 & 0.72 (0.05) & 6.88 (0.03) & 0.86 (0.07) & & 1.00 (0.05) & 6.80 (0.02) & 0.98 (0.06) & 6.62 (0.52) & 13.77(0.23) & 18.82(1.41)  & 1.69(1.46) \\
78 & 0 & G174.50-19.88 & 174.507/-19.887 & \nodata & \nodata & \nodata &  & 0.37 (0.05) & 7.75 (0.07) & 0.93 (0.16) & 2.32 (0.51) & 9.64(0.27) & 10.61(1.69)  & 1.00(0.36) \\
79 & 0 & G174.57-13.66 & 174.573/-13.670 & 1.53 (0.08) & 5.37 (0.01) & 0.39 (0.02) & & 2.39 (0.08) & 5.37 (0.01) & 0.36 (0.02) & 5.95 (0.32) & 11.53(0.49) & 13.42(3.45)  & 2.42(2.42) \\
79 & 1 & G174.57-13.66 & 174.573/-13.670 & 1.45 (0.05) & 6.27 (0.02) & 0.85 (0.04) & & 1.67 (0.05) & 6.26 (0.01) & 0.89 (0.04) & 10.19 (0.53) & 9.69(0.93) & 9.55(1.62)  & 2.42(2.42) \\
80 & 0 & G174.70-15.48 & 174.704/-15.482 & 1.46 (0.05) & 5.83 (0.02) & 0.88 (0.04) & & 2.41 (0.06) & 5.85 (0.01) & 0.82 (0.02) & 13.22 (0.47) & 14.05(0.23) & 22.63(1.11)  & 12.01(3.26) \\
81 & 0 & G174.81-15.15 & 174.814/-15.153 & 1.35 (0.06) & 5.69 (0.01) & 0.58 (0.03) & & 2.19 (0.07) & 5.72 (0.01) & 0.53 (0.02) & 7.79 (0.37) & 6.95(0.10) & 0.00(0.00)  & 2.76(0.74) \\
82 & 0 & G174.88-17.11 & 174.880/-17.114 & 0.43 (0.03) & 5.84 (0.08) & 2.05 (0.18) & & 0.66 (0.04) & 5.92 (0.05) & 1.77 (0.12) & 7.91 (0.68) & 11.30(0.26) & 11.95(1.04)  & 2.96(0.97) \\
83 & 0 & G175.31-20.50 & 175.320/-20.503 & \nodata & \nodata & \nodata &  & 0.39 (0.08) & 8.42 (0.04) & 0.41 (0.10) & 1.10 (0.34) & 9.15(0.28) & 5.50(0.69)  & 0.69(0.07) \\
84 & 0 & G175.34-10.82 & 175.342/-10.826 & 0.78 (0.09) & 5.62 (0.02) & 0.29 (0.04) & & 0.97 (0.10) & 5.54 (0.01) & 0.30 (0.04) & 2.04 (0.35) & 7.71(0.37) & 0.00(0.00)  & 0.72(0.18) \\
84 & 1 & G175.34-10.82 & 175.342/-10.826 & 0.32 (0.06) & 6.40 (0.07) & 0.72 (0.17) & & 0.54 (0.04) & 6.37 (0.07) & 1.25 (0.17) & 4.66 (0.74) &  \nodata & \nodata & 0.72(0.18) \\
85 & 0 & G175.49-16.80 & 175.495/-16.802 & 1.91 (0.09) & 5.56 (0.01) & 0.29 (0.02) & & 3.16 (0.08) & 5.55 (0.01) & 0.42 (0.01) & 8.98 (0.36) & 9.61(0.18) & 9.36(1.01)  & 5.31(1.75) \\
85 & 1 & G175.49-16.80 & 175.495/-16.802 & 0.33 (0.04) & 5.95 (0.11) & 2.11 (0.25) & & \nodata & \nodata & \nodata & \nodata &    \nodata & \nodata & 5.31(1.75) \\
86 & 0 & G175.97-20.38 & 175.979/-20.384 & \nodata & \nodata & \nodata &  & 0.42 (0.07) & 7.50 (0.03) & 0.39 (0.08) & 1.13 (0.30) & 8.96(0.23) & 2.89(0.26)  & 0.47(0.13) \\
87 & 0 & G176.52-09.80 & 176.528/-9.802 & 0.95 (0.06) & 7.41 (0.02) & 0.64 (0.05) & & 1.94 (0.06) & 7.38 (0.01) & 0.49 (0.02) & 6.55 (0.33) & 6.61(0.19) & 0.00(0.00)  & 2.03(0.57) \\
88 & 0 & G177.97-09.72 & 177.978/-9.727 & 2.49 (0.07) & 7.15 (0.01) & 0.47 (0.01) & & 3.96 (0.07) & 7.15 (0.00) & 0.45 (0.01) & 12.52 (0.35) & 14.85(0.26) & 12.43(0.47)  & 8.86(1.58) \\
89 & 0 & G178.72-07.01 & 178.725/-7.012 & 1.10 (0.06) & 7.02 (0.01) & 0.38 (0.02) & & 1.81 (0.06) & 7.03 (0.01) & 0.38 (0.02) & 4.96 (0.27) & 12.50(0.32) & 8.39(0.64)  & 3.12(3.00) \\
90 & 0 & G178.98-06.74 & 178.989/-6.749 & 1.52 (0.06) & 7.75 (0.01) & 0.66 (0.03) & & 2.43 (0.06) & 7.72 (0.01) & 0.63 (0.02) & 11.00 (0.43) & 11.70(0.27) & 13.49(0.92)  & 6.68(1.73) \\
91 & 0 & G181.42-03.73 & 181.428/-3.733 & \nodata & \nodata & \nodata &  & 0.39 (0.04) & -5.39 (0.07) & 1.41 (0.16) & 4.09 (0.61) & 11.60(0.16) & 7.98(0.45)  & 3.16(0.67) \\
92 & 0 & G181.84+00.31 & 181.846/0.317 & \nodata & \nodata & \nodata &  & 0.30 (0.05) & -9.05 (0.11) & 1.44 (0.26) & 3.11 (0.73) & 7.57(0.16) & 1.89(0.43)  & 3.58(1.39) \\
92 & 1 & G181.84+00.31 & 181.846/0.317 & \nodata & \nodata & \nodata &  & 0.83 (0.05) & 3.09 (0.04) & 1.47 (0.09) & 8.79 (0.74) & 9.90(0.14) & 10.23(0.44)  & 3.58(1.39) \\
93 & 0 & G185.33-02.12 & 185.339/-2.127 & 0.40 (0.03) & -0.35 (0.10) & 1.54 (0.22) & & 0.69 (0.07) & -0.32 (0.15) & 1.64 (0.22) & 8.34 (1.39) & 7.52(0.32) & 2.06(0.29)  & 2.21(0.91) \\
93 & 1 & G185.33-02.12 & 185.339/-2.127 & 0.38 (0.03) & 1.52 (0.11) & 1.52 (0.26) & & 0.55 (0.06) & 1.35 (0.18) & 1.69 (0.34) & 6.90 (1.55) & 7.89(0.20) & 9.34(0.61)  & 2.21(0.91) \\
93 & 2 & G185.33-02.12 & 185.339/-2.127 & 0.19 (0.04) & 3.75 (0.10) & 1.05 (0.24) & & 0.23 (0.03) & 3.92 (0.12) & 1.57 (0.30) & 2.68 (0.63) &  \nodata & \nodata & 2.21(0.91) \\
94 & 0 & G192.12-10.90 & 192.129/-10.902 & 1.15 (0.06) & 10.04 (0.02) & 0.65 (0.04) & & 1.69 (0.06) & 10.06 (0.01) & 0.75 (0.03) & 8.75 (0.45) & 17.91(0.14) & 18.54(0.41)  & 3.26(1.22) \\
95 & 0 & G192.28-11.33 & 192.283/-11.339 & 0.98 (0.05) & 10.16 (0.02) & 1.08 (0.06) & & 1.52 (0.04) & 10.06 (0.02) & 1.25 (0.04) & 13.21 (0.54) & 16.76(0.26) & 28.67(0.62)  & 1.48(1.13) \\
96 & 0 & G198.03-15.24 & 198.039/-15.249 & 0.50 (0.07) & -0.08 (0.05) & 0.66 (0.11) & & 0.78 (0.08) & -0.15 (0.03) & 0.63 (0.07) & 3.37 (0.51) & 13.07(0.39) & 9.79(0.40)  & 1.21(0.42) \\
98 & 0 & G038.95-00.47 & 38.96/-0.47 & -1.49 (0.03) & 41.55 (0.04) & 3.82 (0.09) & & -0.79 (0.02) & 41.78 (0.10) & 7.24 (0.23) & 70.05 (2.96) & 20.22(0.13) & 102.24(1.44)  & 53.90(5.39) \\
98 & 1 & G038.95-00.47 & 38.96/-0.47 & \nodata & \nodata & \nodata &  & -0.16 (0.06) & 61.30 (0.19) & 1.09 (0.45) & 2.05 (1.13) & 7.86(0.11) & 8.62(0.59)  & 53.90(5.39) \\
98 & 2 & G038.95-00.47 & 38.96/-0.47 & 0.34 (0.02) & 80.14 (0.20) & 5.77 (0.48) & & 0.54 (0.03) & 78.92 (0.19) & 5.31 (0.47) & 48.74 (4.94) &  \nodata & \nodata & 53.90(5.39) \\
98 & 3 & G038.95-00.47 & 38.96/-0.47 & \nodata & \nodata & \nodata &  & 0.24 (0.04) & 83.97 (0.29) & 3.02 (0.65) & 12.47 (3.34) & 6.05(0.13) & 3.11(0.67)  & 53.90(5.39) \\
99 & 0 & G043.02+08.36 & 43.02/8.37 & 0.28 (0.03) & 4.00 (0.05) & 0.87 (0.11) & & 0.36 (0.03) & 4.04 (0.04) & 1.09 (0.10) & 3.33 (0.42) & 11.55(0.17) & 12.62(0.64)  & 5.77(1.69) \\
99 & 1 & G043.02+08.36 & 43.02/8.37 & 0.17 (0.03) & 8.32 (0.10) & 1.19 (0.23) & & \nodata & \nodata & \nodata & \nodata &    \nodata & \nodata & 5.77(1.69) \\
99 & 2 & G043.02+08.36 & 43.02/8.37 & 0.16 (0.03) & 12.62 (0.07) & 0.72 (0.18) & & \nodata & \nodata & \nodata & \nodata &    \nodata & \nodata & 5.77(1.69) \\
100 & 0 & G046.75-07.68 & 46.76/-7.69 & 0.48 (0.06) & 7.30 (0.03) & 0.56 (0.07) & & 0.98 (0.07) & 7.21 (0.02) & 0.51 (0.04) & 4.15 (0.43) & 11.18(0.22) & 4.89(0.24)  & 0.65(0.32) \\
101 & 0 & G047.74-05.56 & 47.75/-5.57 & 0.50 (0.06) & 8.23 (0.03) & 0.52 (0.07) & & 0.62 (0.06) & 8.18 (0.03) & 0.62 (0.07) & 3.29 (0.48) & 11.68(0.21) & 5.17(0.29)  & 1.92(0.64) \\
102 & 0 & G048.25-05.73 & 48.25/-5.74 & 1.14 (0.07) & 9.01 (0.01) & 0.43 (0.03) & & 1.70 (0.06) & 9.01 (0.01) & 0.52 (0.02) & 7.38 (0.37) & 11.61(0.25) & 12.68(0.88)  & 3.97(1.25) \\
103 & 0 & G048.40-05.82 & 48.41/-5.83 & 1.30 (0.07) & 9.19 (0.01) & 0.54 (0.03) & & 2.10 (0.06) & 9.25 (0.01) & 0.60 (0.02) & 10.54 (0.48) & 9.55(0.22) & 8.31(0.69)  & 4.73(1.77) \\
104 & 0 & G048.65-00.29 & 48.66/-0.29 & -0.43 (0.06) & 5.55 (0.06) & 0.84 (0.13) & & -0.50 (0.04) & 5.60 (0.09) & 2.53 (0.21) & 10.73 (1.18) &  \nodata & \nodata & 26.80(2.68) \\
104 & 1 & G048.65-00.29 & 48.66/-0.29 & -0.53 (0.03) & 33.11 (0.11) & 4.56 (0.26) & & -0.51 (0.02) & 33.37 (0.14) & 6.25 (0.33) & 27.22 (1.88) & 11.21(0.09) & 23.54(0.58)  & 26.80(2.68) \\
105 & 0 & G048.82-03.82 & 48.82/-3.83 & \nodata & \nodata & \nodata &  & 0.32 (0.05) & 9.31 (0.05) & 0.57 (0.11) & 1.57 (0.41) & 10.58(0.23) & 4.37(0.31)  & 0.41(0.20) \\
106 & 0 & G049.06-04.18 & 49.06/-4.18 & \nodata & \nodata & \nodata &  & 0.29 (0.04) & 9.76 (0.11) & 1.52 (0.26) & 3.63 (0.81) & 10.41(0.16) & 5.14(0.28)  & 0.63(0.23) \\
107 & 0 & G049.76-07.25 & 49.77/-7.26 & 0.52 (0.05) & 21.31 (0.04) & 0.85 (0.09) & & 0.91 (0.05) & 21.26 (0.02) & 0.81 (0.05) & 5.90 (0.48) & 8.53(0.21) & 6.47(0.46)  & 0.46(0.05) \\
108 & 0 & G052.99+03.07 & 53.00/3.08 & \nodata & \nodata & \nodata &  & 0.25 (0.04) & 6.98 (0.13) & 1.50 (0.31) & 3.21 (0.87) & 4.20(0.27) & 0.56(0.23)  & 2.21(0.52) \\
108 & 1 & G052.99+03.07 & 53.00/3.08 & 0.50 (0.04) & 10.33 (0.05) & 1.33 (0.12) & & 0.64 (0.05) & 10.26 (0.04) & 1.12 (0.10) & 6.16 (0.75) & 9.97(0.18) & 11.25(0.58)  & 2.21(0.52) \\
112 & 0 & G054.03-02.38 & 54.03/-2.39 & 0.60 (0.05) & 17.06 (0.03) & 0.66 (0.07) & & 0.91 (0.05) & 17.13 (0.02) & 0.85 (0.05) & 6.79 (0.57) & 7.83(0.20) & 4.98(0.45)  & 1.24(0.29) \\
113 & 0 & G056.84+04.81 & 56.84/4.82 & 0.50 (0.04) & 10.89 (0.05) & 1.07 (0.11) & & 0.67 (0.04) & 10.94 (0.04) & 1.32 (0.09) & 6.76 (0.64) & 12.46(0.27) & 9.13(0.43)  & 1.10(0.37) \\
114 & 0 & G057.08+04.46 & 57.08/4.46 & 0.67 (0.05) & 10.66 (0.03) & 0.70 (0.07) & & 0.84 (0.05) & 10.57 (0.03) & 0.93 (0.06) & 6.08 (0.55) & 10.80(0.26) & 7.13(0.31)  & 0.91(0.28) \\
115 & 0 & G057.10+03.65 & 57.11/3.66 & 1.02 (0.04) & 11.21 (0.03) & 1.48 (0.06) & & 1.54 (0.04) & 11.19 (0.02) & 1.62 (0.04) & 19.95 (0.71) & 12.84(0.20) & 26.54(1.00)  & 5.67(1.17) \\
116 & 0 & G057.17+03.41 & 57.17/3.42 & 0.54 (0.04) & 10.79 (0.05) & 1.43 (0.12) & & 0.21 (0.03) & 7.23 (0.17) & 2.32 (0.40) & 3.84 (0.87) &  \nodata & \nodata & 2.21(0.96) \\
116 & 1 & G057.17+03.41 & 57.17/3.42 & \nodata & \nodata & \nodata &  & 1.11 (0.04) & 10.94 (0.02) & 1.48 (0.06) & 13.26 (0.69) & 12.11(0.19) & 13.80(0.52)  & 2.21(0.96) \\
117 & 0 & G057.26+04.01 & 57.26/4.01 & 0.79 (0.05) & 10.70 (0.03) & 0.78 (0.06) & & 1.31 (0.05) & 10.74 (0.02) & 0.77 (0.04) & 7.94 (0.50) & 12.33(0.17) & 8.39(0.38)  & 2.07(1.09) \\
118 & 0 & G058.02+03.02 & 58.03/3.02 & 1.11 (0.05) & 9.78 (0.02) & 0.88 (0.04) & & 1.62 (0.06) & 9.83 (0.01) & 0.70 (0.03) & 9.27 (0.50) & 9.77(0.17) & 7.17(0.45)  & 2.87(0.94) \\
119 & 0 & G058.07+03.20 & 58.07/3.21 & 0.89 (0.04) & 9.83 (0.03) & 1.17 (0.07) & & 1.61 (0.05) & 9.81 (0.01) & 0.96 (0.03) & 12.47 (0.59) & 7.40(0.15) & 9.38(1.00)  & 2.23(1.08) \\
120 & 0 & G058.16+03.50 & 58.16/3.51 & 0.46 (0.03) & 10.44 (0.07) & 1.85 (0.16) & & 0.63 (0.03) & 10.45 (0.05) & 2.04 (0.13) & 10.28 (0.84) & 8.71(0.19) & 11.63(0.55)  & 6.26(1.19) \\
121 & 0 & G058.97-01.66 & 58.97/-1.66 & 0.21 (0.04) & 8.42 (0.13) & 1.35 (0.32) & & 0.22 (0.02) & 9.58 (0.18) & 3.74 (0.43) & 6.83 (1.03) & 7.27(0.18) & 1.57(0.28)  & 1.69(0.75) \\
121 & 1 & G058.97-01.66 & 58.97/-1.66 & \nodata & \nodata & \nodata &  & 0.46 (0.03) & 23.09 (0.05) & 1.49 (0.12) & 5.84 (0.64) & 14.01(0.15) & 15.49(0.67)  & 1.69(0.75) \\
122 & 0 & G060.75-01.23 & 60.75/-1.23 & 0.30 (0.03) & 11.26 (0.11) & 2.15 (0.26) & & 0.63 (0.03) & 11.32 (0.05) & 2.13 (0.12) & 12.14 (0.87) & 10.30(0.18) & 23.02(1.89)  & 2.79(1.49) \\
123 & 0 & G061.76-10.80 & 61.77/-10.81 & \nodata & \nodata & \nodata &  & 0.41 (0.05) & 12.08 (0.05) & 0.98 (0.13) & 2.81 (0.48) & 9.04(0.23) & 3.08(0.46)  & 0.27(0.07) \\
124 & 0 & G065.43-03.15 & 65.43/-3.15 & 0.85 (0.04) & 5.91 (0.03) & 1.39 (0.08) & & 0.95 (0.04) & 5.96 (0.02) & 1.09 (0.06) & 8.42 (0.59) & 7.49(0.14) & 8.69(0.55)  & 2.23(0.64) \\
125 & 0 & G070.44-01.54 & 70.44/-1.55 & 0.49 (0.03) & 10.84 (0.06) & 1.65 (0.13) & & 1.03 (0.03) & 10.88 (0.04) & 2.92 (0.09) & 26.72 (1.14) & 16.42(0.12) & 45.20(0.69)  & 7.44(2.01) \\
126 & 0 & G070.72-00.63 & 70.73/-0.63 & 0.44 (0.03) & 10.40 (0.17) & 3.01 (0.41) & & 0.82 (0.02) & 10.21 (0.12) & 4.51 (0.27) & 32.90 (2.17) & 12.58(0.24) & 23.94(2.83)  & 13.70(7.08) \\
126 & 1 & G070.72-00.63 & 70.73/-0.63 & 0.38 (0.06) & 12.90 (0.10) & 1.34 (0.25) & & 0.76 (0.06) & 13.26 (0.04) & 1.54 (0.13) & 10.42 (1.20) & 10.95(0.52) & 16.08(2.17)  & 13.70(7.08) \\
127 & 0 & G037.49+03.03 & 37.49/3.03 & 0.31 (0.03) & 15.86 (0.05) & 1.06 (0.12) & & 0.97 (0.03) & 15.47 (0.02) & 0.95 (0.04) & 9.92 (0.53) & 9.12(0.14) & 10.52(0.47)  & 1.48(0.70) \\
129 & 0 & G062.16-02.92 & 62.17/-2.93 & 0.38 (0.03) & 13.05 (0.08) & 2.12 (0.19) & & 0.73 (0.04) & 12.47 (0.03) & 1.20 (0.08) & 6.96 (0.59) & 7.99(0.11) & 2.49(0.13)  & 0.53(0.30) \\
130 & 0 & G069.57-01.74 & 69.71/-1.70 & \nodata & \nodata & \nodata &  & 0.25 (0.04) & 14.41 (0.10) & 1.41 (0.26) & 3.13 (0.76) & 7.23(0.25) & 2.64(0.24)  & 5.49(0.55) \\
132 & 0 & SDC033.107-0.065 & 33.107/-0.065 & -1.02 (1.04) & 10.61 (1.02) & 2.04 (2.43) & & -1.53 (0.14) & 10.29 (0.09) & 1.87 (0.20) & 37.15 (5.36) & 7.29(0.64) & 2.82(0.68)  & -0.00(-0.00) \\
133 & 0 & SDC033.332-0.531 & 33.332/-0.531 & -0.27 (0.03) & 94.69 (0.21) & 4.40 (0.50) & & 0.29 (0.03) & 91.39 (0.17) & 3.29 (0.40) & 15.42 (2.47) & 13.48(0.15) & 39.56(0.82)  & -0.00(-0.00) \\
134 & 0 & SDC033.382+0.204 & 33.382/+0.204 & -0.69 (0.02) & 11.66 (0.12) & 6.66 (0.28) & & -0.89 (0.03) & 11.62 (0.09) & 6.46 (0.21) & 74.89 (3.26) & 8.16(0.12) & 4.07(0.47)  & -0.00(-0.00) \\
134 & 1 & SDC033.382+0.204 & 33.382/+0.204 & \nodata & \nodata & \nodata &  & -0.33 (0.04) & 55.94 (0.17) & 3.09 (0.39) & 13.33 (2.24) & 6.18(0.16) & 1.21(0.29)  & -0.00(-0.00) \\
135 & 0 & SDC033.568+0.027 & 33.568/+0.027 & -0.34 (0.04) & 9.59 (0.25) & 3.74 (0.63) & & -0.50 (0.04) & 9.90 (0.25) & 5.26 (0.42) & 34.90 (3.98) &  \nodata & \nodata & -0.00(-0.00) \\
135 & 1 & SDC033.568+0.027 & 33.568/+0.027 & -0.86 (0.08) & 12.18 (0.04) & 0.89 (0.11) & & -0.96 (0.08) & 11.73 (0.04) & 1.28 (0.13) & 16.27 (2.07) & 4.71(0.41) & 2.08(0.39)  & -0.00(-0.00) \\
135 & 2 & SDC033.568+0.027 & 33.568/+0.027 & \nodata & \nodata & \nodata &  & -0.26 (0.04) & 54.92 (0.25) & 3.65 (0.61) & 12.69 (2.77) &  \nodata & \nodata & -0.00(-0.00) \\
135 & 3 & SDC033.568+0.027 & 33.568/+0.027 & \nodata & \nodata & \nodata &  & 0.28 (0.04) & 71.95 (0.20) & 2.54 (0.47) & 12.82 (3.15) & 9.35(0.10) & 11.76(0.46)  & -0.00(-0.00) \\
135 & 4 & SDC033.568+0.027 & 33.568/+0.027 & -0.27 (0.03) & 106.98 (0.27) & 4.90 (0.64) & & -0.45 (0.04) & 102.42 (0.12) & 2.64 (0.30) & 15.86 (2.35) & 12.13(0.08) & 15.64(1.02)  & -0.00(-0.00) \\
136 & 0 & SDC033.622-0.032 & 33.622/-0.032 & -0.49 (0.04) & 11.05 (0.14) & 3.93 (0.34) & & -1.11 (0.04) & 10.96 (0.05) & 2.88 (0.13) & 42.19 (2.49) & 8.00(0.16) & 5.01(0.34)  & -0.00(-0.00) \\
136 & 1 & SDC033.622-0.032 & 33.622/-0.032 & \nodata & \nodata & \nodata &  & 0.42 (0.05) & 34.20 (0.11) & 1.77 (0.27) & 13.45 (2.66) & 11.18(0.15) & 20.63(0.62)  & -0.00(-0.00) \\
136 & 2 & SDC033.622-0.032 & 33.622/-0.032 & -0.90 (0.03) & 107.98 (0.09) & 5.44 (0.22) & & \nodata & \nodata & \nodata & \nodata &   14.09(0.11) & 62.14(1.58)  & -0.00(-0.00) \\
137 & 0 & SDC033.708+0.206 & 33.708/+0.206 & -0.39 (0.05) & 12.45 (0.08) & 1.14 (0.18) & & -0.28 (0.04) & 11.97 (0.14) & 1.87 (0.33) & 7.02 (1.64) & 7.50(0.46) & 1.23(0.59)  & -0.00(-0.00) \\
137 & 1 & SDC033.708+0.206 & 33.708/+0.206 & -0.40 (0.05) & 56.88 (0.08) & 1.52 (0.20) & & -0.83 (0.03) & 56.58 (0.07) & 3.77 (0.16) & 41.30 (2.33) & 10.43(0.45) & 12.93(1.01)  & -0.00(-0.00) \\
138 & 0 & SDC033.743-0.009 & 33.743/-0.009 & -0.52 (0.06) & 11.84 (0.09) & 1.77 (0.22) & & -1.04 (0.05) & 11.37 (0.06) & 2.63 (0.14) & 36.45 (2.49) & 7.19(0.55) & 2.39(0.87)  & -0.00(-0.00) \\
138 & 1 & SDC033.743-0.009 & 33.743/-0.009 & \nodata & \nodata & \nodata &  & -0.56 (0.05) & 56.37 (0.10) & 2.18 (0.23) & 16.40 (2.27) & 5.46(0.46) & 1.16(0.45)  & -0.00(-0.00) \\
138 & 2 & SDC033.743-0.009 & 33.743/-0.009 & -0.66 (0.03) & 110.15 (0.14) & 6.14 (0.33) & & -0.81 (0.03) & 106.26 (0.12) & 7.06 (0.29) & 76.20 (4.12) & 13.74(0.36) & 42.48(2.84)  & -0.00(-0.00) \\
139 & 0 & SDC033.819-0.217 & 33.819/-0.217 & \nodata & \nodata & \nodata &  & -0.55 (0.03) & 10.91 (0.12) & 4.50 (0.28) & 33.54 (2.79) & 9.05(0.50) & 11.39(1.36)  & -0.00(-0.00) \\
140 & 0 & SDC034.056-0.331 & 34.056/-0.331 & \nodata & \nodata & \nodata &  & -0.33 (0.05) & 12.62 (0.10) & 1.45 (0.24) & 6.36 (1.40) & 9.77(0.27) & 17.33(0.91)  & -0.00(-0.00) \\
141 & 0 & SDC034.685-0.729 & 34.685/-0.729 & -0.33 (0.05) & 12.41 (0.13) & 1.95 (0.32) & & -0.84 (0.05) & 12.07 (0.05) & 1.78 (0.12) & 18.26 (1.59) & 9.03(0.12) & 10.13(0.54)  & -0.00(-0.00) \\
141 & 1 & SDC034.685-0.729 & 34.685/-0.729 & -0.72 (0.03) & 44.90 (0.19) & 9.04 (0.34) & & -0.86 (0.03) & 42.23 (0.12) & 6.51 (0.32) & 68.27 (3.90) & 7.20(0.16) & 8.49(0.81)  & -0.00(-0.00) \\
141 & 2 & SDC034.685-0.729 & 34.685/-0.729 & -0.28 (0.05) & 47.50 (0.18) & 2.04 (0.49) & & -0.27 (0.06) & 47.00 (0.15) & 1.68 (0.43) & 5.58 (1.83) & 15.63(0.13) & 10.54(0.65)  & -0.00(-0.00) \\
141 & 3 & SDC034.685-0.729 & 34.685/-0.729 & 1.00 (0.05) & 76.01 (0.04) & 1.59 (0.10) & & 0.52 (0.05) & 73.94 (0.08) & 1.60 (0.18) & 14.24 (2.10) &  \nodata & \nodata & -0.00(-0.00) \\ 
\enddata
\end{deluxetable*}
\end{longrotatetable}

\begin{longrotatetable}
\begin{deluxetable}{llllllll}
\tablecaption{Gaussian fitting parameters of HINSA data.\label{table:HINSA_result}}
\tablewidth{700pt}
\tabletypesize{\scriptsize}
\tablehead{
\colhead{Srcid} &\colhead{Name} & \colhead{gl/gb} & \colhead{N(OH)} & \colhead{N(HINSA)} & \colhead{$p$} &\colhead{Density} &\colhead{$N\rm(H_2)$} \\
 \colhead{} & \colhead{} & \colhead{} & \colhead{$\rm 10^{14} cm^{-2}$} & \colhead{$\rm 10^{18} cm^{-2}$} & \colhead{} & \colhead{$\rm 10^{3} cm^{-3}$} & \colhead{$\rm 10^{21} cm^{-2}$}
}
\startdata
2 & LDN621 & 35.35/2.34 &  4.24( 0.49) & 1.63( 0.55) & 0.81 & \nodata &  4.70( 0.47)  \\ 
4 & CB191 & 37.56/-6.15 &  2.01( 0.33) & 0.38( 0.13) & 0.68 & \nodata &  0.37( 0.04)  \\ 
11 & LDN649 & 43.03/-0.67 &  9.71( 1.65) & 3.20( 1.83) & 0.68 & \nodata & 12.50( 1.25)  \\ 
12 & IREC58 & 43.22/8.33 &  1.26( 0.35) & 4.24( 2.15) & 0.79 & \nodata &  1.07( 0.11)  \\ 
12 & IREC58 & 43.22/8.33 &  1.46( 0.26) & 0.93( 0.36) & 0.72 & \nodata &  1.07( 0.11)  \\ 
14 & LDN658 & 44.38/-2.33 &  1.68( 0.33) & 9.19( 1.19) & 0.63 & \nodata &  1.39( 0.14)  \\ 
20 & LDN666 & 45.25/9.11 &  5.65( 0.47) &10.10( 1.58) & 0.69 & \nodata &  1.96( 0.20)  \\ 
20 & LDN666 & 45.25/9.11 &  1.92( 0.30) & 9.23( 1.70) & 0.61 & \nodata &  1.96( 0.20)  \\ 
26 & MLB10 & 168.18/-16.28 &  8.69( 0.25) & 9.16( 2.64) & 0.90 & \nodata &  4.93( 0.49)  \\ 
29 & LM53 & 171.36/-10.72 &  2.91( 0.14) & 0.74( 0.24) & 0.90 & \nodata &  2.05( 0.20)  \\ 
33 & LDN1525 & 173.3/3.22 &  1.16( 0.21) & 2.33( 0.51) & 0.38 & \nodata &  1.59( 0.16)  \\ 
34 & G171.84-05.22 & 171.848/-5.23 & 11.50( 0.98) & 1.37( 0.74) & 0.84 & 4.21(0.93) & 4.01( 0.80)  \\ 
37 & G175.16-16.74 & 175.166/-16.744 & 21.50( 0.84) & 1.16( 0.56) & 0.58 & 2.00(1.58) & 2.00( 1.56)  \\ 
38 & G175.58-16.60 & 175.583/-16.607 & 18.50( 0.80) & 2.69( 0.96) & 0.58 & 3.18(1.45) & 2.93( 1.23)  \\ 
39 & G176.17-02.10 & 176.177/-2.108 & 18.00( 0.81) & 3.16( 2.08) & 0.56 & 0.29(0.13) & 2.01( 0.80)  \\ 
40 & G176.35+01.92 & 176.352/1.921 & 12.70( 1.03) & 0.47( 0.21) & 0.90 & \nodata &  1.19( 0.34)  \\ 
41 & G176.37-02.05 & 176.374/-2.052 & 17.30( 0.79) & 4.96( 2.59) & 0.56 & 0.22(0.12) & 1.50( 0.77)  \\ 
45 & G178.28-00.61 & 178.286/-0.616 & 18.60( 1.21) & 2.50( 1.05) & 0.72 & \nodata &  2.72( 0.64)  \\ 
45 & G178.28-00.61 & 178.286/-0.616 & 12.20( 1.24) & 1.40( 0.63) & 0.72 & \nodata &  2.72( 0.64)  \\ 
46 & G178.48-06.76 & 178.484/-6.767 & 28.00( 1.24) & 2.67( 1.45) & 0.80 & 2.16(1.08) & 2.55( 1.18)  \\ 
47 & G179.14-06.27 & 179.143/-6.279 & 13.30( 0.72) & 1.89( 0.96) & 0.81 & 1.61(0.89) & 1.44( 0.63)  \\ 
48 & G179.29+04.20 & 179.297/4.2 & 23.00( 1.06) & 2.58( 0.92) & 0.90 & \nodata &  2.89( 1.01)  \\ 
50 & G181.71+04.16 & 181.714/4.163 & 12.40( 0.81) & 1.07( 0.38) & 0.90 & \nodata &  1.95( 1.44)  \\ 
52 & G191.51-00.76 & 191.514/-0.765 & 25.90( 0.98) & 2.73( 1.36) & 0.97 & \nodata &  3.05( 1.33)  \\ 
53 & G201.13+00.31 & 201.138/0.317 & 20.80( 0.77) & 2.05( 1.33) & 0.83 & \nodata &  2.56( 0.26)  \\ 
54 & G201.26+00.46 & 201.269/0.466 & 26.70( 1.26) & 1.16( 0.34) & 0.83 & \nodata &  1.37( 1.56)  \\ 
55 & G201.84+02.81 & 201.841/2.817 &  9.59( 0.65) & 5.67( 2.54) & 0.73 & \nodata &  0.97( 0.52)  \\ 
56 & G158.40-21.86 & 158.401/-21.864 & 10.40( 0.45) & 2.30( 1.19) & 0.64 & 10.95(6.51) & 7.08( 4.11)  \\ 
57 & G158.86-21.60 & 158.862/-21.603 &  9.71( 0.52) & 4.04( 1.88) & 0.64 & 9.07(6.37) & 5.06( 3.48)  \\ 
60 & G168.00-15.69 & 168.003/-15.695 & 10.10( 0.38) & 6.52( 5.18) & 0.73 & 4.89(1.93) & 2.96( 1.11)  \\ 
62 & G168.72-15.48 & 168.728/-15.482 & 15.50( 0.38) & 7.39( 3.93) & 0.73 & 11.53(3.70) & 6.21( 1.83)  \\ 
64 & G169.84-07.61 & 169.849/-7.613 &  5.06( 0.38) & 1.82( 0.96) & 0.85 & 3.78(0.93) & 2.07( 0.45)  \\ 
65 & G170.77-08.51 & 170.771/-8.518 &  4.83( 0.44) & 2.61( 1.88) & 0.76 & 1.58(0.41) & 1.65( 0.40)  \\ 
67 & G171.34-10.67 & 171.343/-10.674 &  8.96( 0.44) & 1.61( 0.72) & 0.71 & 7.09(1.82) & 6.16( 1.46)  \\ 
68 & G173.07-16.52 & 173.079/-16.529 &  8.10( 0.34) &12.70( 9.19) & 0.58 & 3.90(2.52) & 3.51( 2.23)  \\ 
70 & G173.12-13.32 & 173.122/-13.325 &  9.01( 0.41) & 2.24( 0.75) & 0.65 & 2.00(0.96) & 1.83( 0.86)  \\ 
71 & G173.36-16.27 & 173.364/-16.277 & 12.60( 0.40) & 5.65( 3.09) & 0.59 & 10.03(2.12) & 6.89( 1.26)  \\ 
72 & G173.69-15.55 & 173.694/-15.559 &  7.31( 0.38) & 2.14( 1.07) & 0.60 & 6.68(3.11) & 6.26( 2.85)  \\ 
74 & G173.95-13.74 & 173.957/-13.747 & 23.00( 0.65) &14.80( 8.98) & 0.64 & 7.56(3.07) & 6.39( 2.50)  \\ 
76 & G174.39-13.43 & 174.397/-13.44 & 20.70( 0.55) & 7.12( 3.81) & 0.65 & 18.50(17.05) &18.68(17.11)  \\ 
78 & G174.50-19.88 & 174.507/-19.887 &  2.32( 0.51) & 1.64( 1.18) & 0.52 & 1.04(0.39) & 1.00( 0.36)  \\ 
79 & G174.57-13.66 & 174.573/-13.67 &  5.95( 0.32) & 1.89( 0.77) & 0.64 & 2.22(2.24) & 2.42( 2.42)  \\ 
79 & G174.57-13.66 & 174.573/-13.67 & 10.20( 0.53) & 4.07( 1.95) & 0.64 & 2.22(2.24) & 2.42( 2.42)  \\ 
80 & G174.70-15.48 & 174.704/-15.482 & 13.20( 0.47) & 6.55( 4.65) & 0.60 & 13.17(3.80) &12.01( 3.26)  \\ 
81 & G174.81-15.15 & 174.814/-15.153 &  7.79( 0.37) & 0.71( 0.33) & 0.61 & 3.22(0.92) & 2.76( 0.74)  \\ 
83 & G175.31-20.50 & 175.32/-20.503 &  1.10( 0.34) & 0.37( 0.07) & 0.90 & \nodata &  0.69( 0.07)  \\ 
85 & G175.49-16.80 & 175.495/-16.802 &  8.98( 0.35) & 1.33( 0.60) & 0.58 & 6.01(2.07) & 5.31( 1.75)  \\ 
87 & G176.52-09.80 & 176.528/-9.802 &  6.55( 0.33) & 2.63( 1.71) & 0.90 & \nodata &  2.03( 0.57)  \\ 
88 & G177.97-09.72 & 177.978/-9.727 & 12.50( 0.35) & 5.22( 2.14) & 0.73 & 9.98(2.04) & 8.86( 1.58)  \\ 
89 & G178.72-07.01 & 178.725/-7.012 &  4.96( 0.27) & 4.53( 2.64) & 0.80 & 2.20(2.14) & 3.12( 3.00)  \\ 
90 & G178.98-06.74 & 178.989/-6.749 & 11.00( 0.43) & 1.93( 0.96) & 0.80 & 8.45(2.56) & 6.68( 1.73)  \\ 
91 & G181.42-03.73 & 181.428/-3.733 &  4.09( 0.61) & 2.28( 0.86) & 0.90 & \nodata &  3.16( 0.67)  \\ 
92 & G181.84+00.31 & 181.846/0.317 &  3.11( 0.73) & 0.37( 0.12) & 0.90 & \nodata &  3.58( 1.39)  \\ 
92 & G181.84+00.31 & 181.846/0.317 &  8.79( 0.74) & 1.46( 0.74) & 0.90 & \nodata &  3.58( 1.39)  \\ 
95 & G192.28-11.33 & 192.283/-11.339 & 13.20( 0.54) & 4.91( 1.78) & 0.60 & 0.82(0.66) & 1.48( 1.13)  \\ 
100 & G046.75-07.68 & 46.76/-7.69 &  4.15( 0.43) & 1.96( 1.07) & 0.67 & \nodata &  0.65( 0.32)  \\ 
102 & G048.25-05.73 & 48.25/-5.74 &  7.38( 0.36) & 4.09( 1.91) & 0.66 & \nodata &  3.97( 1.25)  \\ 
112 & G054.03-02.38 & 54.03/-2.39 &  6.79( 0.57) & 1.89( 0.97) & 0.74 & \nodata &  1.24( 0.29)  \\ 
113 & G056.84+04.81 & 56.84/4.82 &  6.76( 0.64) & 7.00( 3.32) & 0.63 & \nodata &  1.10( 0.37)  \\ 
115 & G057.10+03.65 & 57.11/3.66 & 19.90( 0.71) & 2.67( 1.25) & 0.72 & \nodata &  5.67( 1.17)  \\ 
118 & G058.02+03.02 & 58.03/3.02 &  9.27( 0.50) & 5.29( 3.67) & 0.80 & \nodata &  2.87( 0.94)  \\ 
119 & G058.07+03.20 & 58.07/3.21 & 12.50( 0.59) & 1.37( 0.65) & 0.78 & \nodata &  2.23( 1.08)  \\ 
122 & G060.75-01.23 & 60.75/-1.23 & 12.10( 0.87) & 0.90( 0.43) & 0.91 & 0.71(0.58) & 2.79( 1.49)  \\ 
126 & G070.72-00.63 & 70.73/-0.63 & 10.40( 1.20) & 1.30( 0.57) & 0.76 & \nodata & 13.70( 7.08)  \\ 
127 & G037.49+03.03 & 37.49/3.03 &  9.92( 0.53) & 4.73( 2.68) & 0.72 & \nodata &  1.48( 0.70)  \\ 
132 & SDC033.107-0.065 & 33.107/-0.065 & 37.20( 5.36) & 0.22( 0.08) & 0.98 & \nodata &  1.07( 0.26)  \\ 
136 & SDC033.622-0.032 & 33.622/-0.032 & 42.20( 2.49) & 0.44( 0.18) & 0.98 & \nodata &  1.90( 0.13)  \\ 
138 & SDC033.743-0.009 & 33.743/-0.009 & 36.40( 2.49) & 1.10( 0.35) & 0.98 & \nodata &  0.91( 0.33)  \\ 
139 & SDC033.819-0.217 & 33.819/-0.217 & 33.50( 2.79) & 0.79( 0.44) & 0.97 & \nodata &  4.33( 0.52)  \\ 
141 & SDC034.685-0.729 & 34.685/-0.729 & 18.30( 1.59) & 0.59( 0.30) & 0.96 & \nodata &  3.85( 0.21)  \\ 
\enddata
\end{deluxetable}
\end{longrotatetable}


\begin{thebibliography}{}
\bibitem[Allen et al.(2012)]{2012AJ....143...97A} Allen, R.~J., Ivette Rodr{\'\i}guez, M., Black, J.~H., et al.\ 2012, \aj, 143, 97
\bibitem[Allen et al.(2015)]{2015AJ....149..123A} Allen, R.~J., Hogg, D.~E., \& Engelke, P.~D.\ 2015, \aj, 149, 123
\bibitem[Barriault et al.(2010)]{2010MNRAS.407.2645B} Barriault, L., Joncas, G., Lockman, F.~J., et al.\ 2010, \mnras, 407, 2645
\bibitem[Bialy et al.(2019)]{2019ApJ...885..109B} Bialy, S., Neufeld, D., Wolfire, M., et al.\ 2019, \apj, 885, 109
\bibitem[Bialy \& Sternberg(2015)]{2015MNRAS.450.4424B} Bialy, S. \& Sternberg, A.\ 2015, \mnras, 450, 4424
\bibitem[Bohlin et al.(1978)]{1978ApJ...224..132B} Bohlin, R.~C., Savage, B.~D., \& Drake, J.~F.\ 1978, \apj, 224, 132
\bibitem[Busch et al.(2019)]{2019ApJ...883..158B} Busch, M.~P., Allen, R.~J., Engelke, P.~D., et al.\ 2019, \apj, 883, 158
\bibitem[Cotten et al.(2012)]{2012AJ....144..163C} Cotten, D.~L., Magnani, L., Wennerstrom, E.~A., et al.\ 2012, \aj, 144, 163
\bibitem[Crutcher(1977)]{1977ApJ...216..308C} Crutcher, R.~M.\ 1977, \apj, 216, 308
\bibitem[Crutcher(1979)]{1979ApJ...234..881C} Crutcher, R.~M.\ 1979, \apj, 234, 881
\bibitem[Crutcher \& Watson(1976)]{1976ApJ...209..778C} Crutcher, R.~M. \& Watson, W.~D.\ 1976, \apj, 209, 778
\bibitem[Crutcher(1973)]{1973ApJ...185..857C} Crutcher, R.~M.\ 1973, \apj, 185, 857
\bibitem[Dawson et al.(2014)]{2014MNRAS.439.1596D} Dawson, J.~R., Walsh, A.~J., Jones, P.~A., et al.\ 2014, \mnras, 439, 1596
\bibitem[Dickey et al.(1981)]{1981A&A....98..271D} Dickey, J.~M., Crovisier, J., \& Kazes, I.\ 1981, \aap, 98, 271
\bibitem[Dickey \& Benson(1982)]{1982AJ.....87..278D} Dickey, J.~M. \& Benson, J.~M.\ 1982, \aj, 87, 278
\bibitem[Dutra \& Bica(2002)]{2002A&A...383..631D} Dutra, C.~M. \& Bica, E.\ 2002, \aap, 383, 631
\bibitem[Ebisawa et al.(2015)]{2015ApJ...815...13E} Ebisawa, Y., Inokuma, H., Sakai, N., et al.\ 2015, \apj, 815, 13
\bibitem[Engelke \& Allen(2018)]{2018ApJ...858...57E} Engelke, P.~D. \& Allen, R.~J.\ 2018, \apj, 858, 57
\bibitem[Engelke \& Allen(2019)]{2019ApJ...874...49E} Engelke, P.~D. \& Allen, R.~J.\ 2019, \apj, 874, 49
\bibitem[Federman et al.(1996)]{1996ApJ...463..181F} Federman, S.~R., Weber, J., \& Lambert, D.~L.\ 1996, \apj, 463, 181
\bibitem[Goldsmith \& Li(2005)]{2005ApJ...622..938G} Goldsmith, P.~F. \& Li, D.\ 2005, \apj, 622, 938
\bibitem[Grossmann et al.(1990)]{1990A&A...240..400G} Grossmann, V., Heithausen, A., Meyerdierks, H., et al.\ 1990, \aap, 240, 400
\bibitem[Harju et al.(2000)]{2000A&A...353.1065H} Harju, J., Winnberg, A., \& Wouterloot, J.~G.~A.\ 2000, \aap, 353, 1065
\bibitem[Haslam et al.(1982)]{1982A&AS...47....1H} Haslam, C.~G.~T., Salter, C.~J., Stoffel, H., et al.\ 1982, \aaps, 47, 1
\bibitem[Heiles(1969)]{1969ApJ...157..123H} Heiles, C.\ 1969, \apj, 157, 123
\bibitem[Heiles et al.(2001)]{2001PASP..113.1247H} Heiles, C., Perillat, P., Nolan, M., et al.\ 2001, \pasp, 113, 1247
\bibitem[Heiles(1968)]{1968ApJ...151..919H} Heiles, C.~E.\ 1968, \apj, 151, 919
\bibitem[Hollenbach et al.(2012)]{2012ApJ...754..105H} Hollenbach, D., Kaufman, M.~J., Neufeld, D., et al.\ 2012, \apj, 754, 105
\bibitem[Kalberla \& Dedes(2008)]{2008A&A...487..951K} Kalberla, P.~M.~W. \& Dedes, L.\ 2008, \aap, 487, 951
\bibitem[Kr{\v{c}}o \& Goldsmith(2010)]{2010ApJ...724.1402K} Kr{\v{c}}o, M. \& Goldsmith, P.~F.\ 2010, \apj, 724, 1402
\bibitem[Le Petit et al.(2016)]{2016A&A...585A.105L} Le Petit, F., Ruaud, M., Bron, E., et al.\ 2016, \aap, 585, A105
\bibitem[Lee et al.(1996)]{1996A&A...311..690L} Lee, H.-H., Herbst, E., Pineau des Forets, G., et al.\ 1996, \aap, 311, 690
\bibitem[Li \& Goldsmith(2003)]{2003ApJ...585..823L} Li, D. \& Goldsmith, P.~F.\ 2003, \apj, 585, 823
\bibitem[Li et al.(2018)]{2018ApJS..235....1L} Li, D., Tang, N., Nguyen, H., et al.\ 2018, \apjs, 235, 1
\bibitem[Liszt \& Lucas(1996)]{1996A&A...314..917L} Liszt, H. \& Lucas, R.\ 1996, \aap, 314, 917
\bibitem[Litvak(1969)]{1969ApJ...156..471L} Litvak, M.~M.\ 1969, \apj, 156, 471
\bibitem[Magnani \& Siskind(1990)]{1990ApJ...359..355M} Magnani, L. \& Siskind, L.\ 1990, \apj, 359, 355
\bibitem[Meng et al.(2013)]{2013ApJS..209...37M} Meng, F., Wu, Y., \& Liu, T.\ 2013, \apjs, 209, 37
\bibitem[Myers et al.(1978)]{1978ApJ...220..864M} Myers, P.~C., Ho, P.~T.~P., Schneps, M.~H., et al.\ 1978, \apj, 220, 864
\bibitem[Negrello et al.(2017)]{2017MNRAS.470.2253N} Negrello, M., Gonzalez-Nuevo, J., De Zotti, G., et al.\ 2017, \mnras, 470, 2253
\bibitem[Nguyen et al.(2018)]{2018ApJ...862...49N} Nguyen, H., Dawson, J.~R., Miville-Desch{\^e}nes, M.-A., et al.\ 2018, \apj, 862, 49
\bibitem[Peek et al.(2018)]{2018ApJS..234....2P} Peek, J.~E.~G., Babler, B.~L., Zheng, Y., et al.\ 2018, \apjs, 234, 2
\bibitem[Peretto et al.(2016)]{2016A&A...590A..72P} Peretto, N., Lenfestey, C., Fuller, G.~A., et al.\ 2016, \aap, 590, A72
\bibitem[Peretto \& Fuller(2009)]{2009A&A...505..405P} Peretto, N. \& Fuller, G.~A.\ 2009, \aap, 505, 405
\bibitem[Planck Collaboration et al.(2011)]{2011A&A...536A..23P} Planck Collaboration, Ade, P.~A.~R., Aghanim, N., et al.\ 2011, \aap, 536, A23
\bibitem[Planck Collaboration et al.(2016)]{2016A&A...594A..28P} Planck Collaboration, Ade, P.~A.~R., Aghanim, N., et al.\ 2016, \aap, 594, A28
\bibitem[Rugel et al.(2018)]{2018A&A...618A.159R} Rugel, M.~R., Beuther, H., Bihr, S., et al.\ 2018, \aap, 618, A159
\bibitem[Sancisi et al.(1974)]{1974A&A....35..445S} Sancisi, R., Goss, W.~M., Anderson, C., et al.\ 1974, \aap, 35, 445
\bibitem[Schultz \& Wiemer(1975)]{1975A&A....43..133S} Schultz, G.~V. \& Wiemer, W.\ 1975, \aap, 43, 133
\bibitem[Schlegel et al.(1998)]{1998ApJ...500..525S} Schlegel, D.~J., Finkbeiner, D.~P., \& Davis, M.\ 1998, \apj, 500, 525
\bibitem[Szymczak \& G{\'e}rard(2004)]{2004A&A...414..235S} Szymczak, M. \& G{\'e}rard, E.\ 2004, \aap, 414, 235
\bibitem[Tang et al.(2020)]{2020RAA....20...77T} Tang, N.-Y., Zuo, P., Li, D., et al.\ 2020, Research in Astronomy and Astrophysics, 20, 077
\bibitem[Tang et al.(2017)]{2017ApJ...839....8T} Tang, N., Li, D., Heiles, C., et al.\ 2017, \apj, 839, 8
\bibitem[Turner(1973)]{1973ApJ...186..357T} Turner, B.~E.\ 1973, \apj, 186, 357
\bibitem[Turner \& Heiles(1974)]{1974ApJ...194..525T} Turner, B.~E. \& Heiles, C.~E.\ 1974, \apj, 194, 525
\bibitem[van der Tak \& van Dishoeck(2000)]{2000A&A...358L..79V} van der Tak, F.~F.~S. \& van Dishoeck, E.~F.\ 2000, \aap, 358, L79
\bibitem[van Dishoeck \& Black(1986)]{1986ApJS...62..109V} van Dishoeck, E.~F. \& Black, J.~H.\ 1986, \apjs, 62, 109
\bibitem[Weinreb et al.(1963)]{1963Natur.200..829W} Weinreb, S., Barrett, A.~H., Meeks, M.~L., et al.\ 1963, \nat, 200, 829
\bibitem[Weselak et al.(2010)]{2010MNRAS.402.1991W} Weselak, T., Galazutdinov, G.~A., Beletsky, Y., et al.\ 2010, \mnras, 402, 1991
\bibitem[Wu et al.(2012)]{2012ApJ...756...76W} Wu, Y., Liu, T., Meng, F., et al.\ 2012, \apj, 756, 76
\bibitem[Xu et al.(2016)]{2016ApJ...819...22X} Xu, D., Li, D., Yue, N., et al.\ 2016, \apj, 819, 22
\bibitem[Zhang et al.(2016)]{2016ApJS..224...43Z} Zhang, T., Wu, Y., Liu, T., et al.\ 2016, \apjs, 224, 43
%
\end{thebibliography}
\end{document}